\documentclass[12pt,preprint]{aastex}

\slugcomment{Published in ApJ}

\shorttitle{H$_2$O formation through the surface reaction OH + H$_2$}
\shortauthors{Oba et al.}

\begin{document}


\title{WATER FORMATION THROUGH A QUANTUM TUNNELING SURFACE REACTION, OH + H$_2$, AT 10 K}


\author{Y. Oba, N. Watanabe, T. Hama, K. Kuwahata, H. Hidaka, and A. Kouchi}
\affil{Institute of Low Temperature Science, Hokkaido University, Sapporo, Hokkaido 060-0819, Japan}
\email{oba@lowtem.hokudai.ac.jp}






\begin{abstract}
The present study experimentally demonstrated that solid H$_{2}$O is formed 
through the surface reaction OH + H$_{2}$ at 10 K. 
This is the first experimental evidence of solid H$_{2}$O formation using hydrogen in its 
molecular form at temperatures as low as 10 K. 
We further found that H$_{2}$O formation through the reaction OH + H$_{2}$ is about one order of 
magnitude more effective than HDO formation through the reaction OH + D$_{2}$. 
This significant isotope effect results from differences in the effective mass of each reaction, indicating that the reactions proceed through quantum tunneling.
\end{abstract}





\section{Introduction}

\subsection{H$_{2}$O formation}

Water (H$_{2}$O) is the most abundant solid component of icy grain mantles in molecular clouds. 
Previous theoretical studies demonstrated that the observed abundance of solid H$_{2}$O in molecular clouds cannot be explained by gas-phase synthesis alone \citep[e.g.][]{hase1992}. 
Therefore, H$_{2}$O is mainly considered to be formed on the surfaces of interstellar grains. 
\citet{tiel1982} proposed a chemical reaction network by which the formation of solid H$_{2}$O begins with hydrogenation of O atoms, O$_{2}$, or ozone, O$_{3}$, on the grain surfaces. 
Since this proposal, many researchers have experimentally explored possible pathways for the formation 
of solid H$_{2}$O under the simulative conditions of dense molecular clouds. 
These studies showed that solid H$_{2}$O is produced through various surface reactions at temperatures as low as 10 K. 
The main surface reactions leading to the formation of solid H$_{2}$O and some related to H$_{2}$O formation 
are listed in Table \ref{tbl-1}.

The simplest pathway toward the formation of solid H$_{2}$O is the 
sequential hydrogenation of O atoms:
\begin{eqnarray}
\label{O+H}{\rm O + H \rightarrow OH,}\\
\label{OH+H}{\rm OH + H \rightarrow H_2O.}
\end{eqnarray}
\citet{cupp2007} proposed that reaction (\ref{OH+H}) is the main route toward H$_{2}$O formation in diffuse and translucent clouds in which H atoms are more prevalent than H$_{2}$ molecules. 
Several groups have experimentally approached these reactions using D atoms instead of H atoms. 
For example, \citet{hira1998} sprayed D atoms over O atoms trapped in an N$_{2}$O matrix at 12 K. 
The O and D atoms were produced by dissociation of, respectively, N$_{2}$O and D$_{2}$ under a dc discharge. 
\citet{duli2010} codeposited O and D atoms produced by dissociating O$_{2}$ and D$_{2}$, respectively, onto non-porous amorphous H$_{2}$O ice at 10 K. 
More recently, \citet{jing2011} deposited D atoms and $^{18}$O (instead of $^{16}$O) at 15--25 K on a synthesized olivine surface. 
All three groups claimed that D$_{2}$O (HDO) formed during their experiments through surface reactions (\ref{O+H}) and (\ref{OH+H}); however, other O-related species, such as O$_{2}$ and O$_{3}$, were also present in or on the samples, and it was difficult to separate information relating to reactions (\ref{O+H}) and (\ref{OH+H}) from the different competing H$_{2}$O-formation channels related to O$_{2}$ and O$_{3}$, shown below. 
Further studies are desired to isolate these reactions (\ref{O+H}) and (\ref{OH+H}) from reactions involving species other than H and O.

The second simplest formation pathway involves the sequential hydrogenation 
of O$_{2}$ via the formation of hydrogen peroxide (H$_{2}$O$_{2})$:
\begin{eqnarray}
\label{O2+H}{\rm O_2 + H \to HO_2,}\\
\label{HO2+H->H2O2}{\rm HO_2 + H \to H_2O_2,}\\
\label{H2O2+H}{\rm H_2O_2 + H \to H_2O + OH.}
\end{eqnarray}
\citet{miya2008} first revealed experimentally that these reactions proceed at 10 K and that a significant isotope effect is apparent for reaction (\ref{H2O2+H}) but not for reactions (\ref{O2+H}) and (\ref{HO2+H->H2O2}). This is consistent with the 
fact that reactions (\ref{O2+H}) and (\ref{HO2+H->H2O2}) have essentially no barrier \citep{walc1988,sell2008}, whereas reaction (\ref{H2O2+H}) has a significant barrier of about 2000 K \citep{kous2006} and requires quantum tunneling \citep{miya2008}.

After the experimental study by \citet{miya2008}, extensive studies were performed on the hydrogenation of O$_{2}$ at low temperatures. 
\citet{iopp2008} revealed that reactions (\ref{O2+H})--(\ref{H2O2+H}) occur even at 28 K. 
Furthermore, it was found that the structure of the H$_{2}$O ice formed through reactions (\ref{O2+H})--(\ref{H2O2+H}) is amorphous \citep{oba2009}, which is consistent with the astronomical observations \citep{smit1989}. 
The recent detection of H$_{2}$O$_{2}$ in the cloud core of $\rho $ Oph A further supported the occurrence of these reactions on the surface of interstellar grains \citep{berg2011} because gas-phase synthesis of 
H$_{2}$O$_{2}$ through reactions (\ref{O2+H}) and (\ref{HO2+H->H2O2}) is not favored \citep{mous2007}. 
In the gas phase, H$_{2}$O$_{2}$ formed through reaction (\ref{HO2+H->H2O2}) further dissociates into two OH species using the excess energy from H$_{2}$O$_{2}$ formation \citep{mous2007}. 
In the solid phase, the release of the excess energy from reaction (\ref{HO2+H->H2O2}) into the cold surface will stabilize the formed H$_{2}$O$_{2}$ on the surface. 
Even though the dissociation occurs, OH radicals are expected to recombine, resulting in the recovery of H$_{2}$O$_{2}$ (reaction (\ref{2OH->H2O2})) and/or the formation of H$_{2}$O (reaction (\ref{2OH->H2O})), as follows: 
\begin{eqnarray}
\label{2OH->H2O2}{\rm OH + OH \to H_2O_2,}\\
\label{2OH->H2O}{\rm OH + OH \to H_2O + O.}
\end{eqnarray}

The reaction HO$_{2}$ + H in the gas phase was theoretically predicted to follow two channels to produce H$_{2}$O via an intermediate species as follows:
\begin{eqnarray}
\label{HO2+H->OD}{\rm HO_2 + H \to H_2OO^* \to H_2O + O(^{1}D),}\\
\label{HO2+H->OP}{\rm HO_2 + H \to H_2O..O \to H_2O + O(^3P),}
\end{eqnarray}
where H$_{2}$OO* and H$_{2}$O..O denote short-lived intermediate species \citep{mous2007}. 
Reaction channel (\ref{HO2+H->OD}) as a whole is slightly exothermic by 10 kJ mol$^{-1}$, whereas the latter part of the reaction (H$_{2}$OO* $\to $ H$_{2}$O + O($^{1}$D)) is highly endothermic by 151 kJ 
mol$^{-1}$ \citep{mous2007}, indicating that reaction channel (\ref{HO2+H->OD}) would not go to completion if the intermediate H$_{2}$OO* dissipated the excess energy to a cold surface. 
Reaction (\ref{HO2+H->OP}) has a huge barrier of about 9300 K, implying that the reaction proceeds only with difficulty, even with tunneling under astrophysically relevant conditions.
Furthermore, the branching ratio to produce O atom is only 2\% for reaction of HO$_2$ + H in the gas phase \citep{keys1986}.
Although \citet{cupp2010} claimed that the reaction HO$_{2}$ + H $\to$ H$_2$O + O occurred in their O$_{2}$/H codeposition experiments, this reaction would be unlikely in the present experiment.

Reactions of HO$_{2}$ with H may also lead to the formation of H$_{2}$ and 
O$_{2}(^{3}\Sigma _{g}^{-})$ \citep{mous2007}:
\begin{eqnarray}
\label{HO2+H->H2+O2}{\rm HO_2 + H \to H_2 + O_2(^{3}\Sigma _{g}^{-}).}
\end{eqnarray}
In the gas phase, this reaction occurs much less frequently than reaction (\ref{HO2+H->H2O2}) \citep{keys1986}. 
Even if O$_{2}$ were reproduced through reaction (\ref{HO2+H->H2+O2}), it must be hydrogenated once again to yield HO$_{2}$ and finally H$_{2}$O$_{2}$ like in the O$_{2}$-hydrogenation experiments.

The sequential hydrogenation of ozone (O$_{3})$ may occur in molecular 
clouds:
\begin{eqnarray}
\label{O3+H}{\rm O_3 + H \to O_{2} + OH.}
\end{eqnarray}
The products (O$_{2}$ and OH) may be further used for H$_{2}$O formation. 
\citet{mokr2009} investigated these reactions at 10 K by irradiating solid O$_{3}$ on H$_{2}$O ice with D atoms. 
The reaction products were analyzed using temperature-programmed desorption (TPD) methods. 
\citet{roma2011} performed experiments of the hydrogenation of O$_{3}$ at higher temperatures (25, 40, and 50 K), and the reaction products were monitored in-situ by Fourier-transform infrared (FTIR) spectroscopy. 
Both studies provided results indicating that the formation of H$_{2}$O was initiated by reaction (\ref{O3+H}).

Note that OH is always produced along each of these three pathways: reaction (\ref{O+H}) for the hydrogenation of O atoms, reaction (\ref{H2O2+H}) for that of O$_{2}$, and reaction (\ref{O3+H}) for that of O$_{3}$. 
Hence, in addition to the H-atom related reactions shown above, the following H$_{2}$-related reaction would lead to H$_{2}$O formation:
\begin{eqnarray}
\label{OH+H2}{\rm OH + H_{2} \to H_{2}O + H, }
\end{eqnarray}
which has an activation barrier of $\sim $2100 K in the gas phase \citep{atki2004}. 
Despite the large activation barrier of reaction (\ref{OH+H2}), reaction (\ref{OH+H2}) is often thought of as the most common route to the formation of H$_{2}$O on grain surfaces in dense molecular clouds in which typical 
temperatures are about 10 K, the UV flux is very low, and H$_{2}$ is the dominant H-bearing species over H atoms \citep{tiel1982,tiel2005,cupp2007}. 
At temperatures as low as 10 K, Arrhenius-type reactions having an activation barrier of $\sim $2000 K occur only rarely. 
Therefore, reaction (\ref{OH+H2}) requires quantum tunneling to proceed on grain surfaces, as observed in the gas-phase reaction at low temperatures \citep{nguy2011}. 
The tunneling reaction is described in further detail in the next section.

Reaction (\ref{OH+H2}) was experimentally studied in the gas phase \citep{ravi1981,olde1992,alag1993,talu1996,kras2004,orki2006}. 
In most previous studies, OH was produced by photolysis of H$_{2}$O or H$_{2}$O$_{2}$, or indirectly 
by photolysis of N$_{2}$O to generate O($^{1}$D) atoms, which further reacted with H$_{2}$O \citep{vagh1989,olde1992,talu1996,orki2006}. 
OH generated through those energetic processes should have significant excess energy that should be 
sufficient to overcome the activation barrier of reaction (\ref{OH+H2}). 
Hence, to investigate whether reaction (\ref{OH+H2}) occurs through quantum tunneling on cold grains, non-energetic OH must be used for experiments. 
In recent experimental studies of low-temperature surface reactions, OH radicals were produced through various reactions and were further used in secondary reactions. 
For example, hydrogenation of O$_{3}$ yielded OH as a reaction product (reaction (\ref{O3+H})); however, OH formed in reaction (\ref{O3+H}) could be highly energetic because the heat of the reaction \citep[$\sim $39,000 K;][]{keys1979} is partitioned into OH \citep{roma2011}, which is not suitable for investigating quantum tunneling reactions.

We recently succeeded in producing cold OH by dissociating H$_{2}$O in a microwave-induced plasma, followed by cooling to 100 K before reaching the reaction substrate \citep{oba2010a,oba2010b,oba2011}. 
A similar technique for producing OH in the electronic and vibrational ground states was applied to crossed beam studies of reaction (\ref{OH+H2}) in the gas phase \citep{alag1993}, where OH was produced in a radiofrequency-induced H$_{2}$O plasma. 
We found that the cold OH can react with CO and HOCO on a substrate even at 10 K to yield CO$_{2}$ \citep{oba2010a} and H$_{2}$CO$_{3}$ \citep{oba2010b}, respectively. 
Moreover, it was experimentally demonstrated that cold OH reacts with another OH to yield H$_{2}$O$_{2}$ and H$_{2}$O through reactions (\ref{2OH->H2O2}) and (\ref{2OH->H2O}), respectively, even at 40 K \citep{oba2011}. 
Consequently, our experimental method is useful for studying chemical reactions related to cold OH on the surfaces of interstellar grain mantles. 

In the present study, we first performed the codeposition of cold OH with H$_{2}$ on a cold substrate to study reaction (\ref{OH+H2}) as a possible route to the formation of H$_{2}$O on the surfaces of interstellar grains in dense molecular clouds. 
Next, OH and/or H$_{2}$ isotopologues, such as OD and D$_{2}$, were studied experimentally in an investigation of the isotope effect on the reactions to decisively conclude that these reactions occur via quantum tunneling.

\subsection{Quantum tunneling reaction}

The thermally activated reaction (\ref{OH+H2}) is not expected to occur on grain 
surfaces at 10 K for the reasons given above. Previous experimental and 
theoretical studies proposed that quantum tunneling plays a significant role 
in reaction (\ref{OH+H2}) in the gas phase, particularly at low temperatures \citep[e.g.][]{talu1996,nguy2011}. \citet{wata2008} described the importance of quantum tunneling in various chemical reactions 
on low-temperature grain surfaces in molecular clouds. 
Under the simple assumption that a reaction occurs through a rectangular barrier of height $E_{\rm a}$ and width $a$ on a potential energy surface, the rate of a quantum tunneling reaction ($k_{\rm q})$ is approximately represented by the following equation:
\begin{eqnarray}
\label{tunneling}{k_{\rm {q}}\approx  v_{0} {\rm exp} [-{(\frac{2a}{\hbar})}{(2\textit{mE}_{a})^{\frac{1}{2}}],}} 
\end{eqnarray}
where $v_{0}$ and $m$ denote the frequency of harmonic motion and particle mass, respectively. 
In the case of a bimolecular atom-transfer reaction represented by A + XB $\to $ AX + B, the center of mass translated from the initial state (A + XB) to the final state (AX + B) of a reaction on the potential energy surface is represented as the effective mass $m_{\rm c}$ which is defined by the following equation:
\begin{eqnarray}
\label{mass}{m_{\rm c} = \frac{m_{\rm a}m_{\rm b}(1 + c)^{2}+m_{\rm b}m_{\rm x}c^{2} + 
m_{\rm a}m_{\rm x}}{M(1 + c^{2})}, }
\end{eqnarray}
where $m_{\rm a}$, $m_{\rm x}$, and $m_{\rm b}$ are the masses of A, X, and B atoms, respectively, $M$ is the mass of the linear triatomic molecule AXB ($M$ = $m_{\rm a}+m_{\rm x}+m_{\rm b})$, and $c$ is the ratio of the infinitesimal change of one bond distance to that of the other bond distance \citep{john1966}. 
When bimolecular atom-transfer reactions A + XB $\to $ AX + B proceed through quantum tunneling, an effective mass $m_{\rm c}$ should be used to indicate the tunneling mass in Equation (\ref{tunneling}) \citep{hida2009}.

Observation of an isotope effect on a reaction can be an indicator of quantum tunneling. 
This is because tunneling effects strongly depend on the particle mass, such that a tunneling reaction involving a species with a lighter effective mass proceeds faster than with a heavier effective mass. 
Previous studies have investigated the isotope effect on other astrophysically relevant chemical reactions around 10 K. 
\citet{hida2007} performed hydrogenation/deuteration of solid CO at 10--20 K and found that hydrogenation of CO proceeds faster than deuteration of CO by a factor of about 12. 
H--D isotope exchange reactions of methanol (CH$_{3}$OH) were also demonstrated to occur through quantum tunneling \citep{naga2007}, which would contribute to the observed abundance of deuterated methanol, such as CHD$_{2}$OH.

The importance of the effective mass on quantum tunneling reactions has been demonstrated experimentally. 
\citet{hida2009} extensively investigated H-D exchange reactions of formaldehyde (H$_{2}$CO) at 10--20 K. The effective mass of the D-atom abstraction by H atoms from D$_{2}$CO is $\sim $1, whereas that of the H-atom abstraction by D atoms from H$_{2}$CO is $\sim $0.5, which affects the rate of each reaction \citep{hida2009}.

Most quantum tunneling surface reactions previously studied were H- and D-atom-related reactions. 
Previous experimental studies of H$_{2}$-related quantum tunneling reactions have investigated the isotope effect on the reactions of methyl radicals (CH$_{3})$ or its deuterated species (e.g. CHD$_{2}$ and CD$_{3})$ in solid parahydrogen ($E_{a}\sim $ 5300 K) at 5 K \citep{momo1998,hosh2004}. 
Although these pioneering studies are important for understanding quantum tunneling reactions related to 
H$_{2}$, their relevance to astrophysics is rather weak because their experiments were conducted under unfeasible conditions (in a parahydrogen matrix at 5 K) in molecular clouds. 

In the present study, we experimentally examined surface reactions related to H$_{2}$ and OH isotopologues, as shown below, to investigate the isotope effects on reaction (\ref{OH+H2}):
\begin{eqnarray}
\label{OD+H2}{\rm OD + H_{2} \to HDO + H,}\\
\label{OH+D2}{\rm OH + D_{2} \to HDO + D,}\\
\label{OD+D2}{\rm OD + D_{2} \to D_{2}O + D,}\\
\label{OH+HD->H2O}{\rm OH + HD \to H_{2}O + D,}\\
\label{OH+HD->HDO}{\rm OH + HD \to HDO + H,}\\
\label{OD+HD->HDO}{\rm OD + HD \to HDO + D,}\\
\label{OD+HD->D2O}{\rm OD + HD \to D_{2}O + H.}
\end{eqnarray}
The use of isotopically substituted hydrogen (HD or D$_{2})$ or hydroxyl radical (OD) in place of H$_{2}$ or OH, respectively, can alter the effective mass of a reaction relative to that of reaction (12). 
In this case, the rate of a quantum tunneling reaction should differ from that of reaction (12) as well. 
Reported values of the activation barrier and the calculated effective mass for each reaction are summarized in Table \ref{tbl-2}. 
According to \citet{hida2009}, the value of $c$ in Equation (\ref{mass}) was set to --1 in the present reaction system. 
An observation that the efficiency of each reaction varied according to the effective mass would strongly indicate 
that these reactions proceed through quantum tunneling. 

\section{EXPERIMENTAL}

\subsection{Apparatus and experimental conditions for the surface reactions}

Experiments were performed using the Apparatus for SUrface Reaction in Astrophysics system. 
Details of this apparatus and experimental conditions have been described previously \citep{wata2006,naga2007,oba2010a,oba2011}. 
Briefly, OH radicals, together with H atoms, were produced by the dissociation of H$_{2}$O in a microwave 
discharge plasma \citep{timm1998,timm1999} in a Pyrex tube. 
H$_{2}$O dissociation can lead directly to the formation of H$_{2}$ and O. 
The species produced in the OH source are denoted ``H$_{2}$O fragments'' in this paper. 
H$_{2}$O fragments were transferred via a series of poly(tetrafluoroethylene) and a cold aluminum (Al) tube, which was connected to a He refrigerator, and they impinged on the substrate. 
Initially, OH radicals may be excited when formed in the plasma, but after many collisions with the inner wall of a cold Al pipe, they are thermalized to 100 K. 
That is, they should be vibrationally and electronically in the ground state before reacting on the substrate, as demonstrated in gas-phase collision experiments \citep{alag1993,kohn2011}. 
This was confirmed experimentally, as described in the next section. 
The deposition rate of OH radicals was not measured directly; the upper limit of the deposition rate should be 1.5 $\times$ 10$^{13}$ radicals cm$^{-2}$ s$^{-1}$, which is equal to the deposition rate of H$_{2}$O when the microwave source is turned off. 
H$_{2}$ molecules delivered from a separate gas line through a capillary plate were codeposited with the H$_{2}$O fragments onto the substrate over 60 minutes. 
The flux of H$_{2}$ was calculated to be 2.0 $\times $ 10$^{14}$ molecules cm$^{-2}$ s$^{-1}$ based on the measured pressure inside the main chamber when filled with only H$_{2}$. 

The reaction products were monitored in-situ by reflection--absorption-type FTIR with a resolution of 4 cm$^{-1}$ in the spectral range 4000--800 cm$^{-1}$. 
The desorbed species from the substrate were monitored using a quadrupole mass spectrometer (QMS). The temperature of the substrate, which was connected to another He refrigerator, was kept at 10 K in each 
experiment. 
The reactions in Table \ref{tbl-2} did not occur if the codeposition experiments were performed at temperatures above 20 K.

After codeposition, the TPD spectra were obtained by the QMS at a heating rate of 4 K minutes$^{-1}$.

In the experiments using H$_{2}$- and OH-isotopologues, the HD and D$_{2}$ fluxes were the same as the H$_{2 }$ flux. 
QMS measurements and a resonance-enhanced multiphoton ionization (REMPI) method (see Section 2.2) confirmed that almost all ($>$95{\%}) D$_{2}$O or H$_{2}$O introduced into the radical source was fragmented when the microwave source was turned on. 
On this basis, the deposition rate of OD was considered to be similar to that of OH because the flow rates of D$_{2}$O and H$_{2}$O into the radical source were adjusted to the same value.

\subsection{In Situ Detection of OH Radicals}

The direct detection of OH from the source was performed using another apparatus, named RASCAL, in our laboratory. RASCAL is described in detail elsewhere \citep{wata2010,hama2011}. 
OH from the radical source was selectively ionized by (2 + 1) REMPI approximately 5 mm from the exit of the aluminum pipe and detected using a time-of-flight mass spectrometer. 

\begin{figure}[p]
\plotone{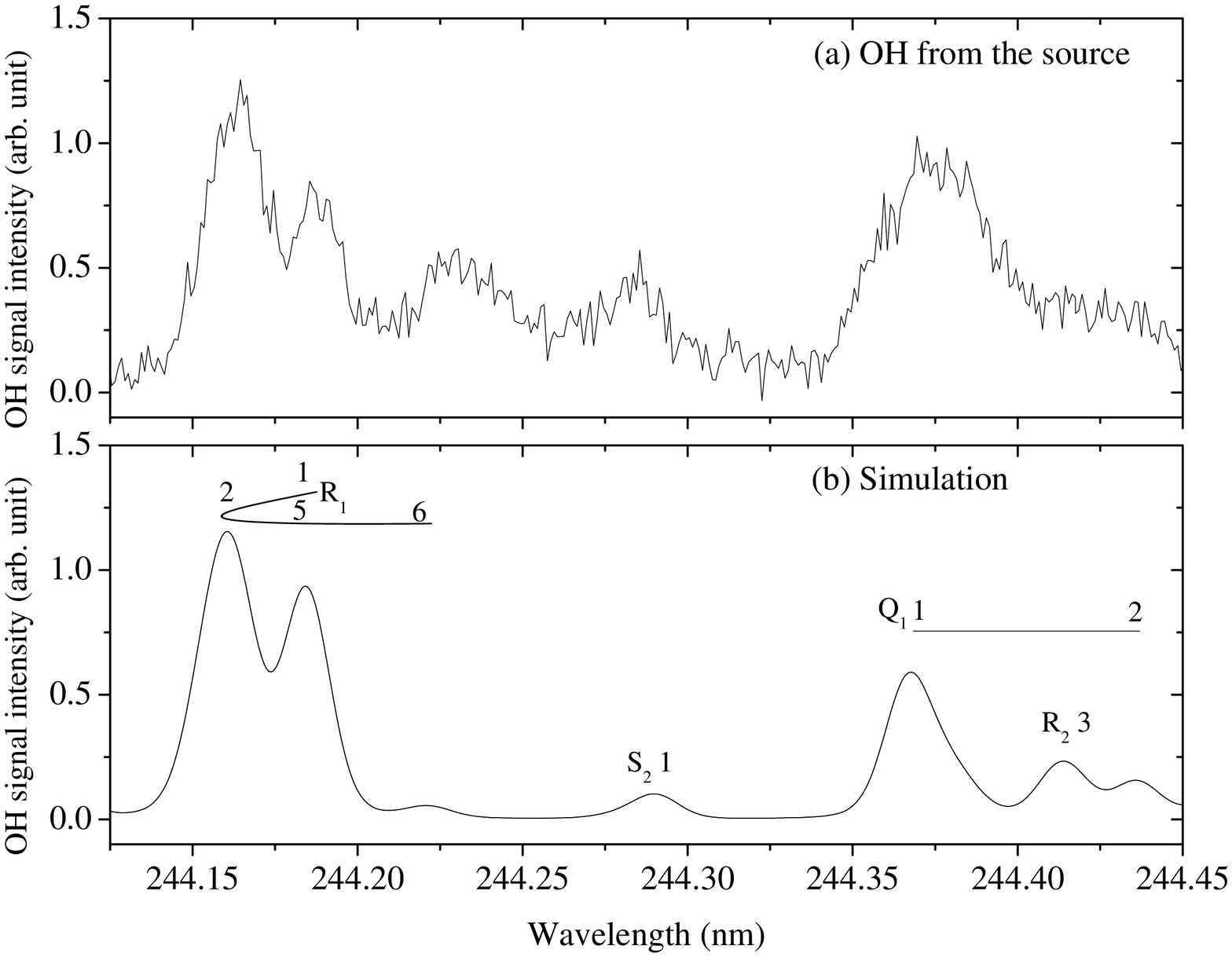} 
  \caption{ (a) (2 + 1) REMPI spectrum of OH emitted from the radical source through the Al pipe at room 
temperature and (b) the best-fit simulated spectrum for $T_{\rm rot}$ = 300 K. 
The capital numbers 1--6 indicate rotational levels of Q, R, and S branches associated with the $X^{2}\Pi $ ground state of OH. The subscript numbers 1 
and 2 represent the lower or upper members of the rotational level in the $X^{2}\Pi $ ground state, respectively, because spin-orbit coupling causes 
the rotational levels to split into two components \citep{debe1991,diek1962}.}
  \label{fig1}
\end{figure}

Figure \ref{fig1}(a) shows the REMPI spectrum of OH($v$=0) via the $D^{2}\Sigma^{-}(v'$=0)$\leftarrow X^{2}\Pi (v$=0) transition under conditions in which the Al pipe was equilibrated at room temperature. 
The presence of vibrationally excited OH($v$=1) was sought using the REMPI, with 3$^{2}\Sigma ^{-}$($\it v'$=0)$\leftarrow X^{2}\Pi (v$=1) transition; however, no such evidence was obtained. 
\citet{gree2005} calculated that the two-photon cross-sections of the $D^{2}\Sigma^{-}(v'=0)\leftarrow X^{2}\Pi (v$=0) and 3$^{2}\Sigma^{-}($v'$=0)\leftarrow X^{2}\Pi (v$=1) transitions are comparable; therefore, the present results indicate that OH from the radical source was electronically and vibrationally in the ground state. 
The REMPI spectrum of OH($v$=0) was assigned and modeled using the PGOPHER simulation program to determine the rotational structure (C. M. Western, available from http://pgopher.chm.bris.ac.uk). 
The constants associated with the relevant electronic states were compared with those in the literature \citep{diek1962,gree2005,hube1979}. 
Figure \ref{fig1}(b) shows the results of a spectral simulation of OH($v$=0) assuming a rotational temperature of 300 K. 
By comparison with the spectral simulation, the OH($v$=0) from the radical source was expected to be almost 
thermally equilibrated with the Al pipe before reacting on the substrate.

In an additional step, the adsorbates were photodesorbed using the weak second harmonic radiation (532 nm) of a Nd:YAG laser and were analyzed by the REMPI method. 
We detected OH($v$=0) signals in the species photodesorbed from the cold substrate (10--40 K) during the deposition of H$_{2}$O fragments. 
Photolysis of H$_{2}$O or H$_{2}$O$_{2}$ on the substrate, which could be potential sources for OH($v$=0), was negligible at 532 nm \citep{chu2005,koba1983}; therefore, this result clearly indicates the presence of OH in the electronic and vibrational ground states on the substrate at 10 K. 
The rotational temperature of the photodesorbed OH($v$=0) was about 100 K. 
This value is slightly higher than the substrate temperature. 
The OH radical is considered to gain a small amount of energy during photostimulated desorption. 
The present results are consistent with previous experimental results obtained using a similar method at 3.5 K \citep{zins2011}.

\section{RESULTS AND DISCUSSION}

\subsection{OH + H$_{2}$}

\begin{figure}
\epsscale{0.8}
 \plotone{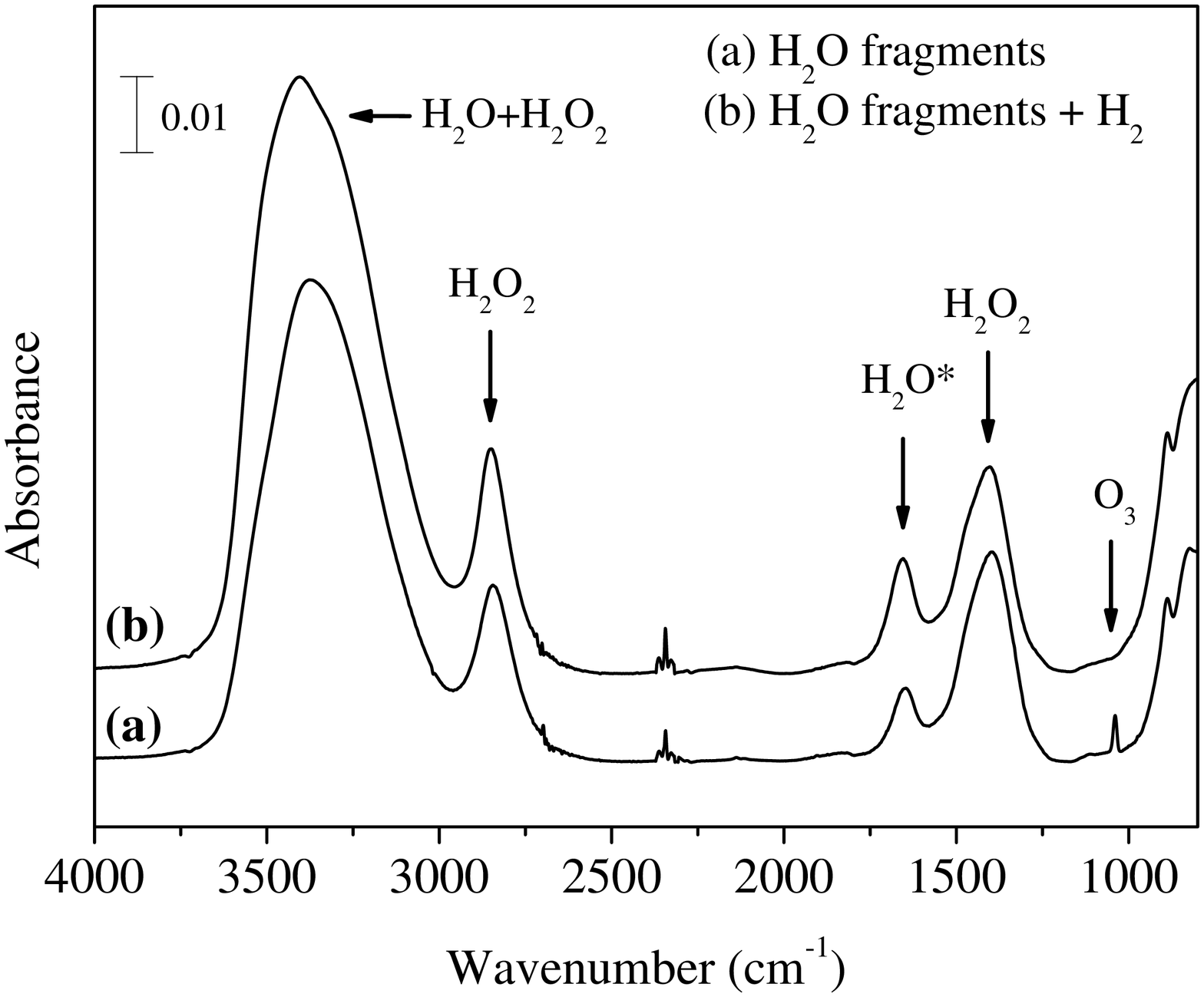}
 \plotone{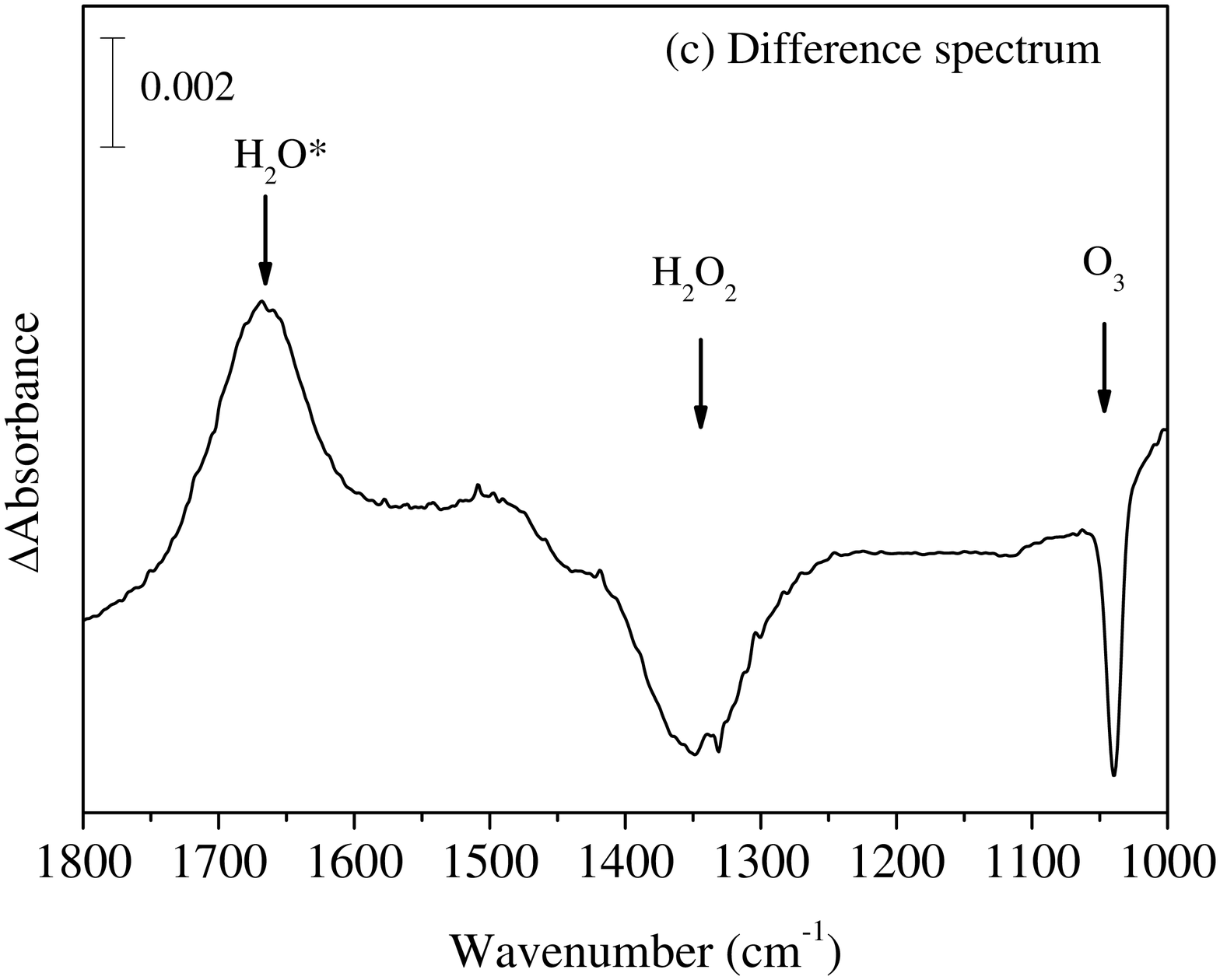}
  \caption{ FTIR spectra after (a) deposition of H$_{2}$O fragments and (b) codeposition 
of H$_{2}$O fragments with H$_{2}$. (c) The spectral differences relative to 
the H$_{2}$O-fragment deposition sample. The peak with the asterisk (*) was 
used for quantification purposes. The peaks at $\sim$2300 cm$^{-1}$ and $\sim$2700 cm$^{-1}$ were derived from the background CO$_2$ and the inherent noises caused by vibration of the He refrigerator, respectively.}
  \label{fig2}
\end{figure}

The deposition only of H$_{2}$O fragments on a substrate at 10 K resulted in the formation of H$_{2}$O, H$_{2}$O$_{2}$, and O$_{3}$, as identified in the product IR spectrum at 1650, 1400, and 1040 cm$^{-1}$, respectively (Figure \ref{fig2}(a)).
Hereafter, this experiment is denoted as the H$_2$O-fragment deposition. 
In contrast, these molecules were not observed in the H$_{2}$O-fragment deposition at 60 K \citep{oba2011}. 
If H$_{2}$O, H$_{2}$O$_{2}$, and O$_{3}$ formed in the gas phase, they should be adsorbed on the substrate even at 60 K, which is well below their desorption temperatures. 
We therefore concluded that these molecules were formed through surface reactions at 10 K. 
The formation pathways of these molecules are described in our recent paper \citep{oba2011}. 
The peak area for each reaction product increased linearly with the deposition time. 
In addition to the three types of molecules listed above, QMS measurements revealed that H$_{2}$ and O$_{2}$, which are infrared-inactive, are present in the gases from the source and/or are produced on the surface.

Figure \ref{fig2}(b) shows the IR spectrum of the reaction product obtained after codeposition of H$_{2}$O fragments with H$_{2}$. 
This spectrum was obtained under the same experimental conditions as those for the H$_2$O-fragment deposition: the deposition rate of OH, the experimental time, and substrate temperatures. 
The resulting spectrum was similar to that obtained in the H$_2$O-fragment deposition, although the peak area associated with the OH bending of H$_{2}$O (1650 cm$^{-1})$ increased by a factor of two compared to that 
observed in the H$_2$O-fragment deposition (Figure \ref{fig2}(c)). 
This result indicates that the introduced H$_{2}$ was used for the formation of H$_{2}$O through surface reactions with O-related species, such as OH, O, O$_{2}$, and/or O$_{3}$. 
We concluded that H$_{2}$O was formed by reaction of H$_{2}$ with OH (reaction (\ref{OH+H2})) on the surface at 10 K, despite the relatively large barrier $E_{\rm a}$ of about 2100 K \citep{atki2004}, for the following 
reasons.

First, we confirmed that codeposition of H$_{2}$ with cold O atoms produced by the dissociation of O$_{2}$ in a microwave-induced plasma did not result in a reaction at 10 K, except that O$_{3}$ formed (Y. Oba et al., unpublished data). 
This result indicates that cold O atoms in an H$_{2}$O plasma will not react with H$_{2}$ either. 
This is in good agreement with the fact that the reaction, 
\begin{eqnarray}
\label{O+H2}{\rm{ O + H_{2} \to OH + H,}}
\end{eqnarray}
is endothermic by 8 kJ mol$^{-1}$ (= 960 K) with a significant activation barrier of 3160 K \citep{baul1992}, although theoretical studies assumed H$_{2}$O formation begins with reaction (22) \citep{caza2010,caza2011}. 
Second, O$_{2}$ never reacts with H$_{2}$ under the present experimental conditions. 
Third, O$_{3}$ does not react with H$_{2}$, as experimentally demonstrated previously \citep{mokr2009,roma2011}. 
In addition, if HO$_{2}$ is produced through reactions of the H$_{2}$O fragments, it may react with H$_{2}$ to yield H$_{2}$O$_{2}$:
\begin{eqnarray}
\label{HO2+H2}{\rm HO_{2} + H_{2} \to H_{2}O_{2} + H.} 
\end{eqnarray}
In this case, the H$_{2}$O$_{2}$ produced may further yield H$_{2}$O through reaction (\ref{H2O2+H}). 
This, however, is unlikely in the present experiment because reaction (\ref{HO2+H2}) is highly endothermic with a considerable activation barrier (Table \ref{tbl-1}).

The column density of the H$_{2}$O product was calculated from the peak area of the OH bending band at 1650 cm$^{-1}$ and the previously reported integrated band strengths, as described in \citet{hida2007}. 
The band strength used was 1.2 $\times $ 10$^{-17}$ cm molecules$^{-1}$ which is by a transmission method \citep{gera1995}.
We confirmed that the ratio of the band strengths for OH stretching and bending bands of H$_2$O is almost independent of the method (transmission or reflection).
While, the difference in the absolute value of band strengths for the reflection and the transmission methods was found to be within the factor of two in our experimental setup \citep{hida2009}.
Moreover, it is difficult to determine the exact column density even by the transmission method in the layered and the mixed samples.
However, since our discussion is based on relative reaction yields, the difference in the absolute value of the band strengths between the two methods little alters the present experimental results and discussion. 

By subtracting the H$_{2}$O band obtained in the H$_{2}$O-fragment deposition from that obtained in the codeposition experiment, we determined the increase in the column density of H$_{2}$O through reaction (12) to be 3.5 $\times $ 10$^{15}$ molecules cm$^{-2}$.
Note that it is not obvious whether H$_2$O yield obtained by the subtraction gives the total yield of H$_2$O by the reaction OH + H$_2$.
In the H$_2$O-fragment deposition, two competing reactions produce H$_2$O: reactions (\ref{2OH->H2O}) and (\ref{OH+H2}).
In the codeposition experiment, additional H$_2$ molecules enhance reaction (\ref{OH+H2}).
That is, a part of OH amount which was used by reaction (\ref{2OH->H2O}) in the H$_2$O-fragment deposition may be consumed by reaction (\ref{OH+H2}). 
Therefore, H$_2$O column density obtained by the subtraction may be underestimated compared to the total yield of H$_2$O by reaction (\ref{OH+H2}) in the codeposition.
Further details on this issue are shown in the next section.

Accompanied by the increase in the H$_{2}$O column density, the peak area of the OH bending in H$_{2}$O$_{2}$ at 1400 cm$^{-1}$ decreased relative to the peak area measured in the H$_{2}$O-fragment deposition (Figure \ref{fig2}(c)). 
This is probably because the abundant H$_{2}$ which was codeposited with H$_{2}$O fragments inhibited the formation of H$_{2}$O$_{2}$ through reaction (\ref{2OH->H2O2}) (OH + OH $\to $ H$_{2}$O$_{2})$; the OH fragments instead formed H$_{2}$O via reaction (\ref{OH+H2}). 
Reaction (\ref{OH+H}) (OH + H $\to $ H$_{2}$O) was assumed to contribute to H$_{2}$O formation only to a minor extent in the present study because most H atoms in the H$_{2}$O fragments must be consumed by H--H 
recombination on the surface or few H atoms were present in the gases from the source. 
In fact, the codeposition of O$_{2}$ with the H$_{2}$O fragments did not increase the amount of H$_{2}$O$_{2 }$ relative to the H$_{2}$O-fragment deposition, despite the barrierless mechanism for O$_{2}$ hydrogenation.

The absorption at 1040 cm$^{-1}$ due to O$_{3}$ was not observed in the codeposition experiment (Figure \ref{fig2}(b)). 
It is possible that O$_{3}$ was consumed by H atoms (reaction (\ref{O3+H})) that formed through reaction (\ref{OH+H2}) and/or the O atoms were similarly consumed by H atoms (reaction (\ref{O+H})) prior to the formation of O$_{3}$ (O$_{2}$ + O $\to $ O$_{3})$. 
Reactions (\ref{O+H}) and (\ref{O3+H}) yield energetic OH, which may further react with H$_{2}$ to yield H$_{2}$O through reaction (\ref{OH+H2}). 
Even if this were the case, the low O$_{3}$ column density produced in the H$_{2}$O-fragment deposition (0.6 $\times $ 10$^{15}$ molecules cm$^{-2})$ suggested that such reactions did not significantly contribute to the observed H$_{2}$O column density. 
Nevertheless, a lack of O$_{3}$ could provide indirect evidence that the reaction OH + H$_{2}$ occurred. 
In fact, the codeposition of D atoms with H$_{2}$O fragments did not yield O$_{3}$ formation (see Section 3.3).

\subsection{OD + H$_{2}$}

Similar experiments were performed using OD instead of OH. If OD reacts with H$_{2}$, the formation of HDO is expected to occur through the following H-atom abstraction from H$_{2}$ by OD:
\setcounter{equation}{14}
\begin{eqnarray}
 {\rm OD + H_{2} \to HDO + H,}
\end{eqnarray}
which has almost the same activation barrier and effective mass as reaction (\ref{OH+H2}) \citep[][Table 2]{nguy2011}. 
In the OH/H$_{2}$ codeposition experiment described in the previous section, the reaction products were basically the same as those of the H$_2$O-fragment deposition except for the presence of O$_{3}$. 
In contrast, HDO, the reaction product of reaction (\ref{OD+H2}), is not produced unless both H and D-related species are used, that is, HDO is never produced in the deposition only of H$_{2}$O or D$_{2}$O 
fragments. 
The formation of HDO should thus provide direct evidence that reaction (\ref{OD+H2}) occurs at 10 K.

\begin{figure}
\plotone{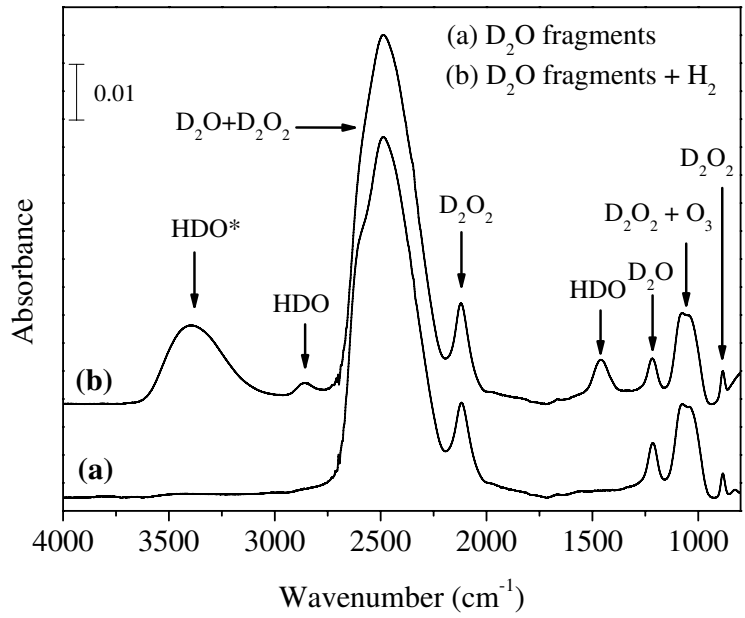} 
  \caption{ FTIR spectra after (a) deposition of D$_{2}$O fragments and (b) codeposition 
of D$_{2}$O fragments with H$_{2}$. The peak with the asterisk (*) was used 
for quantification purposes.}
  \label{fig3}
\end{figure}

Formation of D$_{2}$O, D$_{2}$O$_{2}$, and O$_{3}$ was observed in the product IR spectrum upon the sole deposition at 10 K of D$_{2}$O fragments (Figure \ref{fig3}(a)) (hereafter, this experiment is denoted as the D$_2$O-fragment deposition), which were produced in a manner similar to that used to produce H$_{2}$O fragments. 
These results indicated that the isotopic substitution of H$_{2}$O fragments did not significantly alter the reaction pathways on the cold substrate.

\begin{figure}
\plotone{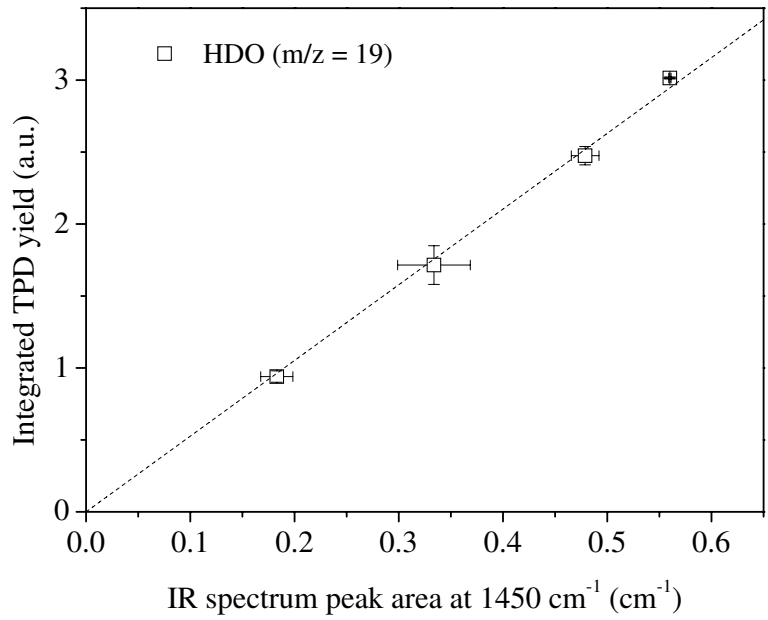} 
  \caption{Integrated TPD yield of HDO ($\it{m/z}$ = 19) as a function of the IR peak area of 
the 1450 cm$^{-1}$ band, which appeared upon the codeposition of D$_{2}$O 
fragments with H$_{2}$. The data points show the averaged values with statistical errors for four different durations. A dashed line represents the best-fitting straight line through the points.}
  \label{fig4}
\end{figure}

The codeposition of D$_{2}$O fragments with H$_{2}$ resulted in the appearance of new peaks at 3392, 2858, and 1453 cm$^{-1}$ (Figure \ref{fig3}(b)). 
Absorption bands at 3392 and 1453 cm$^{-1}$ were typically observed for solid HDO produced by depositing an H$_{2}$O and D$_{2}$O mixture on a substrate at 100 K or lower \citep{horn1958,devl1986,galv2011}. 
The absorption band at 2858 cm$^{-1}$ may correspond to the 2$v_{2}$ vibration of HDO \citep{devl1986}. Because the only H source is H$_{2}$ and other O-related species, such as O atoms and O$_{3}$, hardly react with H$_{2}$, as explained in Section 3.1, we conclude that H$_{2}$ reacted with OD to yield HDO on the surface at 10 K. 
The presence of HDO in the reaction product was also confirmed by the TPD experimental results. 
Figure \ref{fig4} shows the integrated TPD yields of $\it{m/z}$ = 19 (HDO) plotted with the peak area of the band at 1453 cm$^{-1}$.

The column density of HDO was calculated from the peak area of the OH-stretching band, and the previously reported integrated band strength of 1.3 $\times $ 10$^{-16 }$ cm molecule$^{-1}$ for the OH-stretching band of the solid HDO \citep{ikaw1968} was found to be 4.7 $\times $ 10$^{15}$ molecules cm$^{-2}$.

In contrast to the experimental result shown in the previous section, the obtained column density of HDO should represent the total yield of HDO by the reaction OD+ H$_2$ because HDO is not formed in the D$_2$O-fragment deposition.
The HDO yield is very similar to the estimated yield of H$_2$O by the reaction OH + H$_2$ (3.5 $\times $ 10$^{15}$ molecules cm$^{-2}$) in the previous section.
It is reasonably considered that our estimated H$_2$O yield represents the total yield of H$_2$O by the reaction OH + H$_2$.
This is based on an assumption that reaction (\ref{OH+H2}) has the same efficiency with reaction (\ref{OD+H2}) (see Section 3.6 for further details).

\subsection{OH + D$_{2}$}

\begin{figure}
\plotone{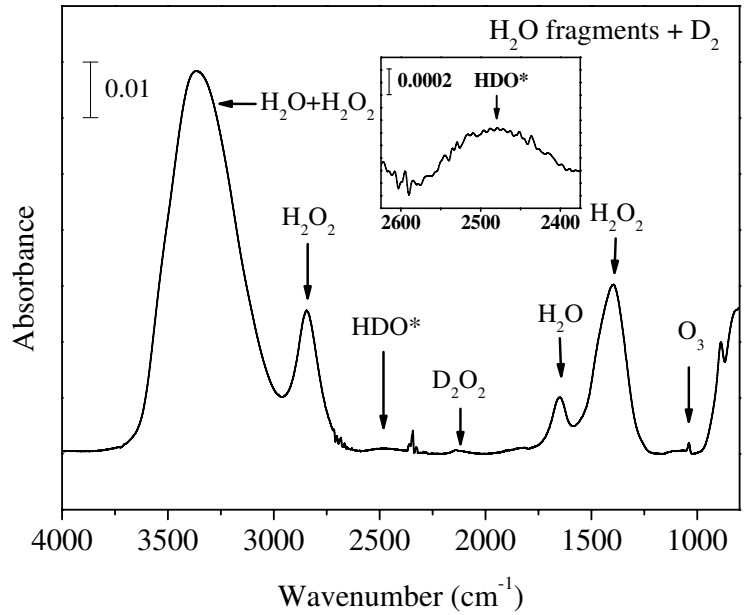} 
  \caption{ FTIR spectrum after codeposition of H$_{2}$O fragments with D$_{2}$. The 
peak with the asterisk (*) was used for quantification purposes. The inset shows the OD-stretching band of HDO at 2475 cm$^{-1}$.}
  \label{fig5}
\end{figure}

We next codeposited H$_{2}$O fragments with D$_{2}$ to investigate the efficiency of reaction (\ref{OH+D2}), which had almost the same barrier as reactions (\ref{OH+H2}) and (\ref{OD+H2}) in the gas phase \citep{nguy2011}. 
The effective mass of reaction (\ref{OH+D2}) was twice that of reactions (\ref{OH+H2}) and (\ref{OD+H2}) (Table \ref{tbl-2}). 
Figure \ref{fig5} shows an IR spectrum of the reaction product obtained after codeposition of H$_{2}$O fragments with D$_{2}$. 
The spectrum resembled that obtained from the H$_2$O-fragment deposition (Figure \ref{fig2}(a)); however, a broad peak grew in at 2475 cm$^{-1}$, which is a typical position for OD-stretching bands. 
In addition, a small peak was observed at 2150 cm$^{-1}$. 
This peak may be associated with D$_{2}$O$_{2 }$ \citep{lann1971}, which is probably produced through 
deuteration of O$_{2}$ by the D atoms formed through reaction (\ref{OH+D2}). 
Because the only D source is D$_{2}$ in this experiment, the appearance of an OD-stretching band indicates that reaction (\ref{OH+D2}) occurred at 10 K. 
The presence of HDO was verified in the TPD experiment.

The column density of the formed HDO was calculated based on the peak area of the OD stretching band at 2475 cm$^{-1}$ and the previously reported integrated band strengths of 5.0 $\times $ 10$^{-17}$ cm molecule$^{-1}$ for the OD stretching band of the solid HDO \citep{ikaw1968}. 
The column density was found to be 2.7 $\times $ 10$^{14}$ molecules cm$^{-2}$, much lower than the column density of the H$_{2}$O formed via reaction (\ref{OH+H2}) under the similar experimental conditions (see Section 3.1). 
A low efficiency was also inferred from the presence of O$_{3}$ in the reaction product (Figure \ref{fig5}). 
As described above, if reaction (\ref{OH+D2}) proceeds efficiently to produce significant quantities of D atoms, O$_{3}$ would be consumed. 
In fact, the codeposition of H$_{2}$O fragments with D atoms did not result in observation of O$_{3}$.

\subsection{OD + D$_{2}$}

\begin{figure}
\plotone{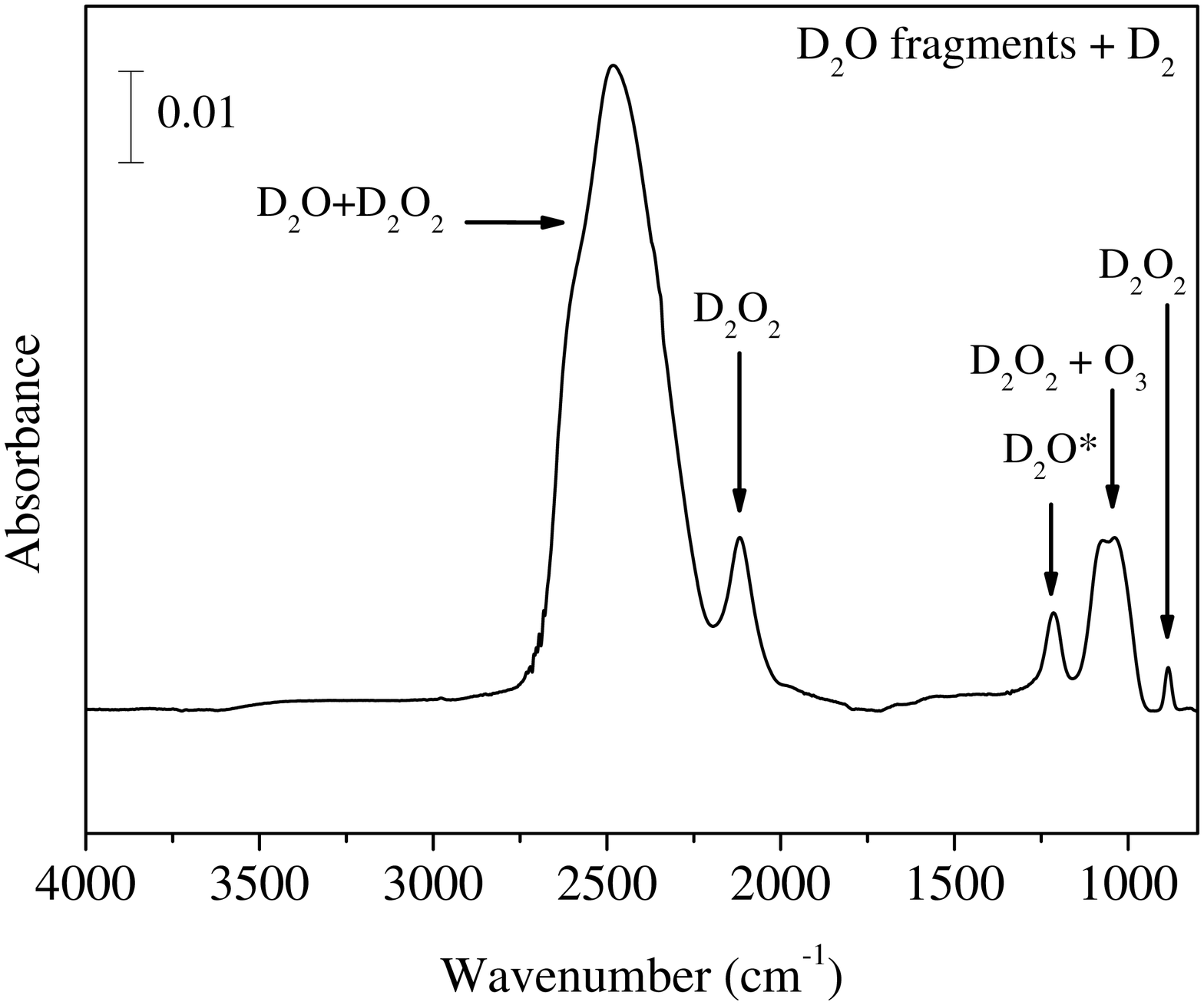} 
  \caption{ FTIR spectrum after codeposition of D$_{2}$O fragments and D$_{2}$. The peak 
with the asterisk (*) was used for quantification purposes.}
  \label{fig6}
\end{figure}

Reactions of OD with D$_{2}$ are expected to yield D$_{2}$O and D atoms as 
follows:
\setcounter{equation}{16}
\begin{eqnarray}
{\rm OD + D_{2} \to D_{2}O + D.}
\end{eqnarray}
Because the effective mass for reaction (\ref{OD+D2}) exceeds those of reactions (\ref{OH+H2}) 
and (\ref{OD+H2}) and is comparable to that of reaction (\ref{OH+D2}), reaction (\ref{OD+D2}) is 
expected to be less efficient than reactions (\ref{OH+H2}) and (\ref{OD+H2}). 
Figure \ref{fig6} shows an IR spectrum obtained after codeposition of D$_{2}$O fragments with 
D$_{2}$. 
The spectrum resembled that of the D$_2$O-fragment deposition (Figure \ref{fig3}(a)). 
A slight increase in the OD bending band of D$_{2}$O was observed at 1200 cm$^{-1}$, which corresponds to an increase in the D$_{2}$O column density by 3.5 $\times $ 10$^{14}$ molecules cm$^{-2}$. 
The band strength used was 8.0 $\times $ 10$^{-18}$ cm molecule$^{-1}$ \citep{miya2008}.

\subsection{OH or OD + HD}

Unlike reactions that involve H$_{2}$ or D$_{2}$, two reaction channels are 
available for OH in the presence of HD:
\begin{eqnarray}
{\rm OH + HD \to H_{2}O + D,}\\
{\rm OH + HD \to HDO + H.}
\end{eqnarray}
These reactions are associated with nearly identical barrier heights \citep{nguy2011} whereas the effective masses differed considerably from one another (0.48 for reaction (\ref{OH+HD->H2O}) and 0.90 for reaction (\ref{OH+HD->HDO})). 
The experimental results shown in Sections 3.1--3.4, therefore, predict that reaction (\ref{OH+HD->H2O}) is more efficient than reaction (\ref{OH+HD->HDO}). 
The branching ratio for these reactions in the gas phase was experimentally estimated, suggesting that reaction (\ref{OH+HD->H2O}) was 3--4 times more efficient than reaction (\ref{OH+HD->HDO}) \citep{talu1996}. 
H$_{2}$O fragments were codeposited with HD to investigate the efficiencies of reactions (\ref{OH+HD->H2O}) and (\ref{OH+HD->HDO}). 

\begin{figure}
\epsscale{0.9}
\plotone{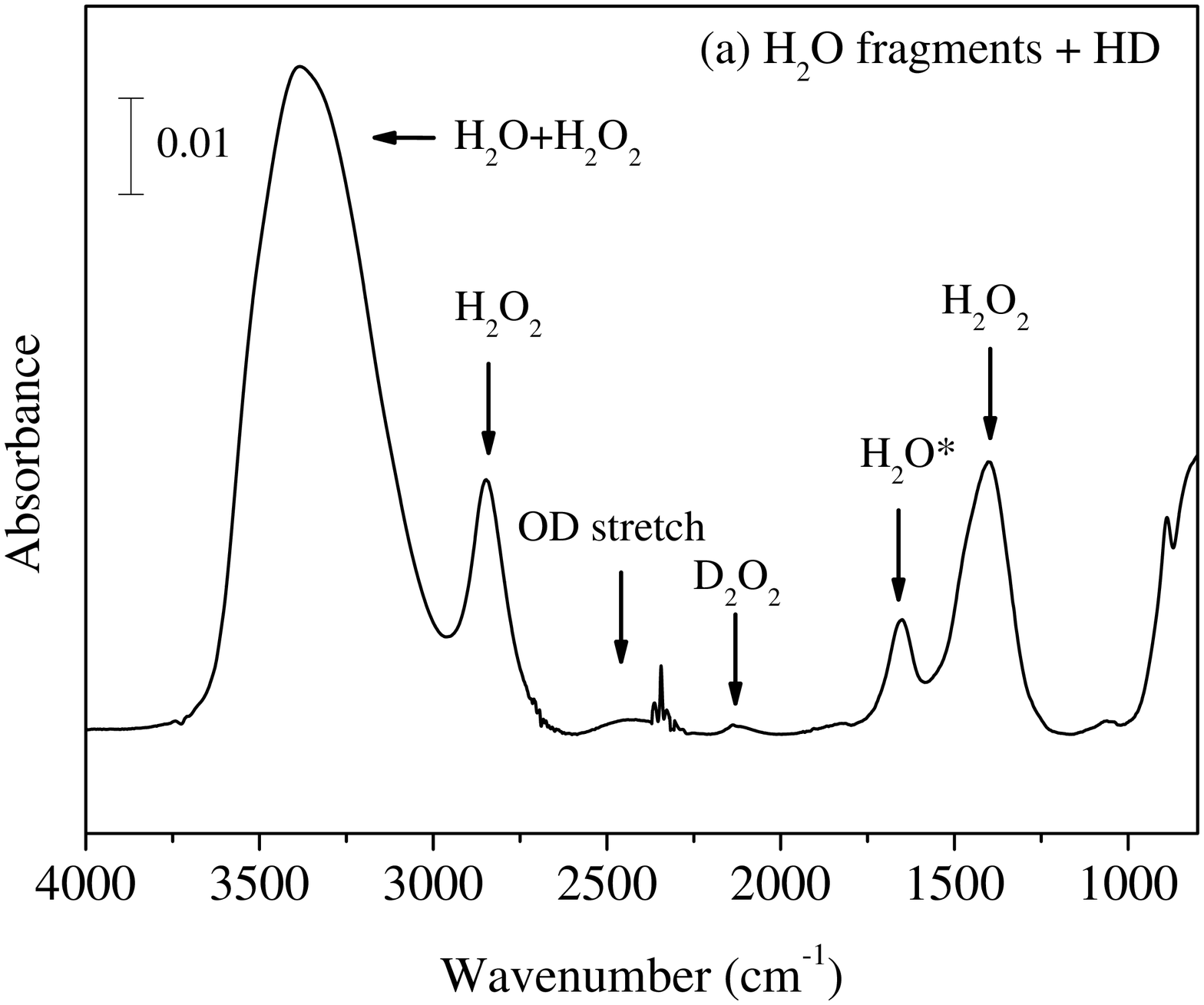} 
\plotone{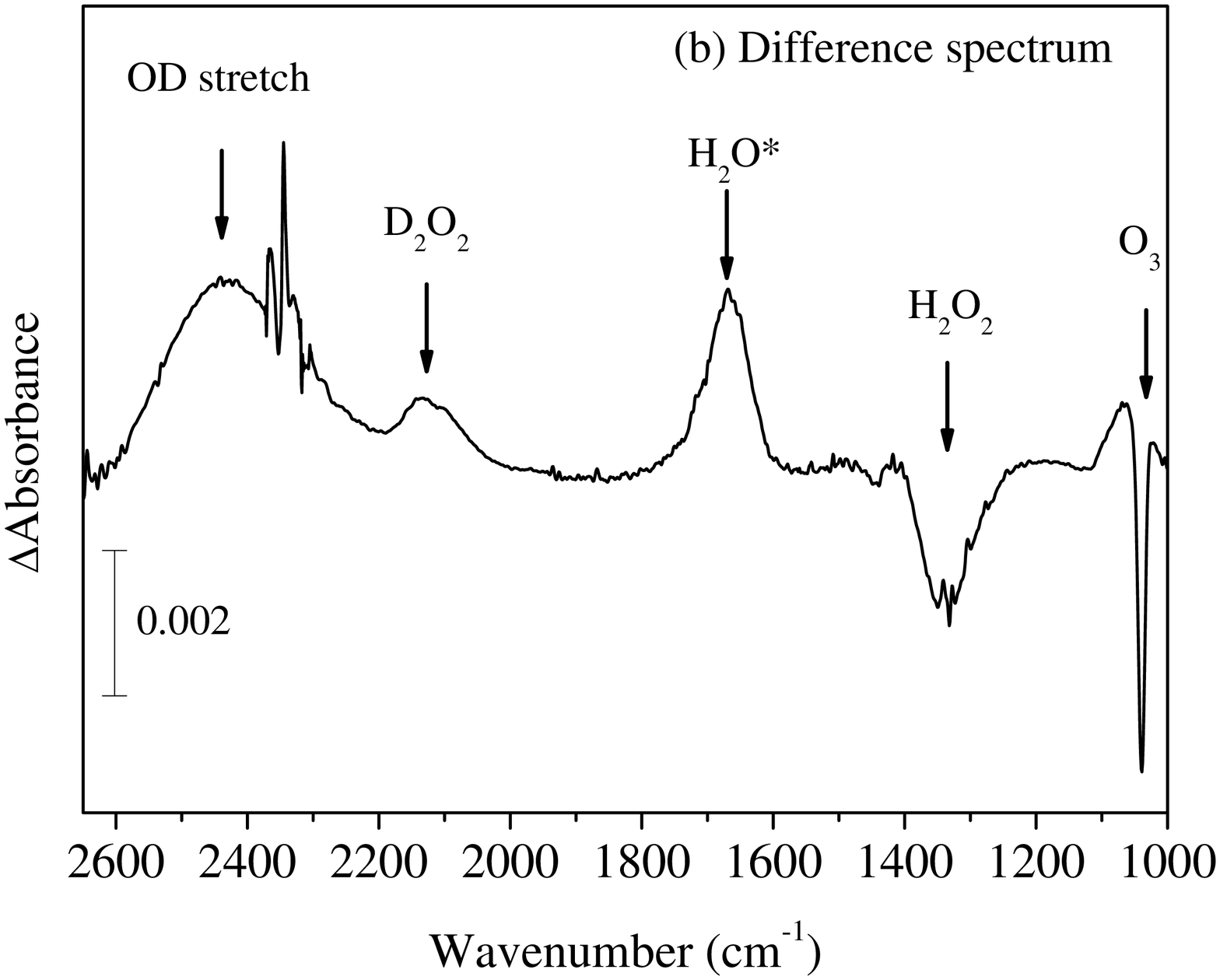}
  \caption{ (a) FTIR spectrum after codeposition of H$_{2}$O fragments with HD and (b) 
the spectral differences relative to the H$_{2}$O-fragment deposition 
sample. The peak with the asterisk (*) was used for quantification purposes. 
The peaks at 2300 cm$^{-1}$ were derived from the background CO$_{2}$.}
  \label{fig7}
\end{figure}

The formation of H$_{2}$O and H$_{2}$O$_{2}$ was observed in the product IR spectrum obtained after codeposition of H$_{2}$O fragments with HD. 
O$_{3}$ was not observed (Figure \ref{fig7}(a)). 
The peak area for H$_{2}$O increased by a factor of 1.5 compared to the H$_2$O-fragment deposition (Figure \ref{fig7}(b)), which corresponded to an increase in the H$_{2}$O column density by 3.1 $\times $ 10$^{15}$ molecules cm$^{-2}$. 
In addition to the H$_{2}$O increase, small peaks were observed at 2435 and 2114 cm$^{-1}$ (Figure \ref{fig7}(b)). 
The former peak corresponded to an OD-stretching band, implying that HDO was formed through the reaction OH + HD; however, the observed OD-stretching band was expected to include large contributions from species other than HDO. 
If reaction (\ref{OH+HD->H2O}) is the dominant channel relative to reaction (\ref{OH+HD->HDO}), D atoms are produced together with H$_{2}$O, which results in the formation of D$_{2}$O$_{2}$ through reactions between D and O$_{2}$ contained in the H$_{2}$O fragments. 
In fact, the peak positions at 2435 and 2114 cm$^{-1}$ were consistent with D$_{2}$O$_{2}$ \citep{lann1971}. 
Assuming that the OD-stretching band was mainly derived from D$_{2}$O$_{2}$, the contributions of HDO to the band should be very small.

\begin{figure}
\plotone{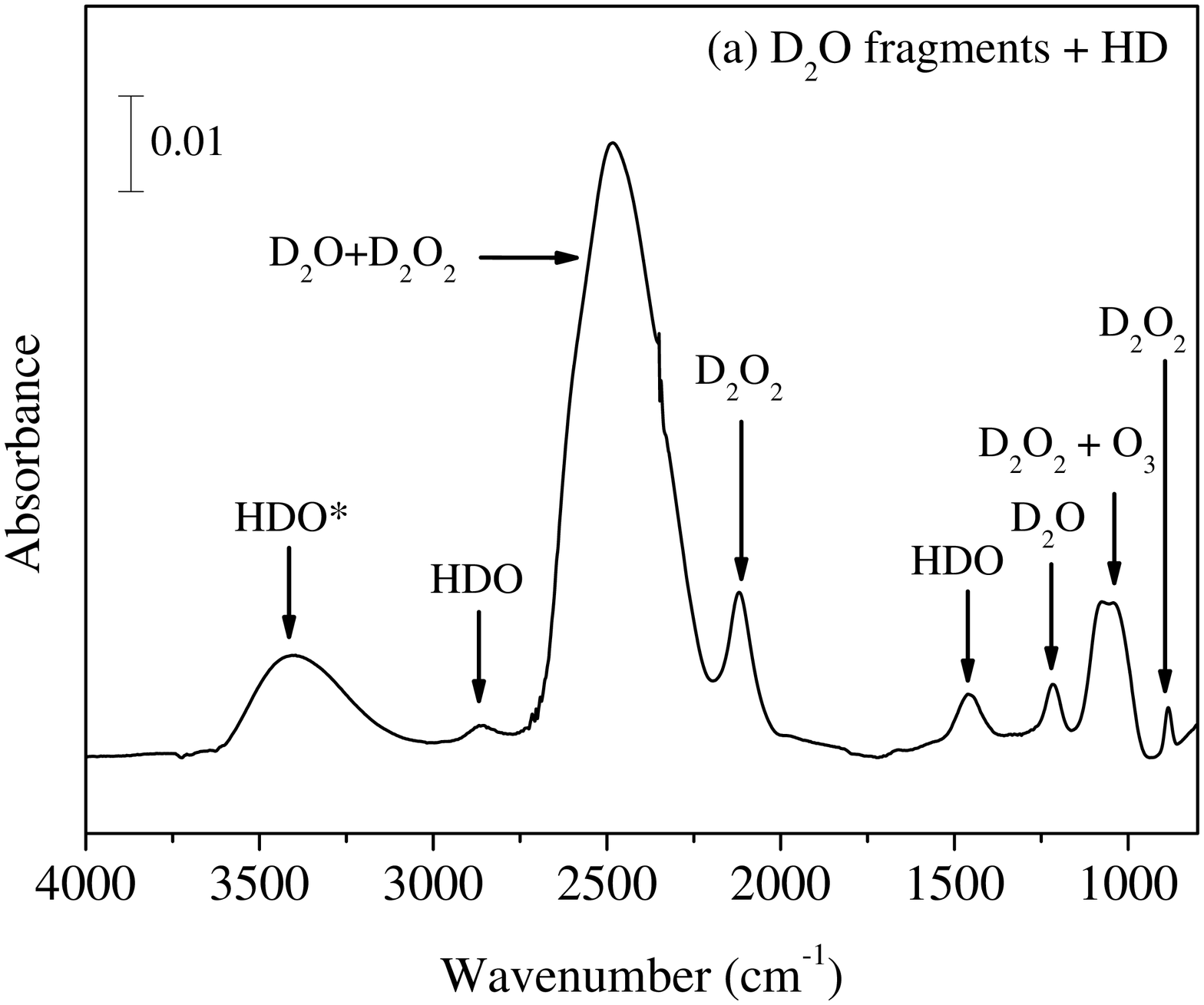} 
\plotone{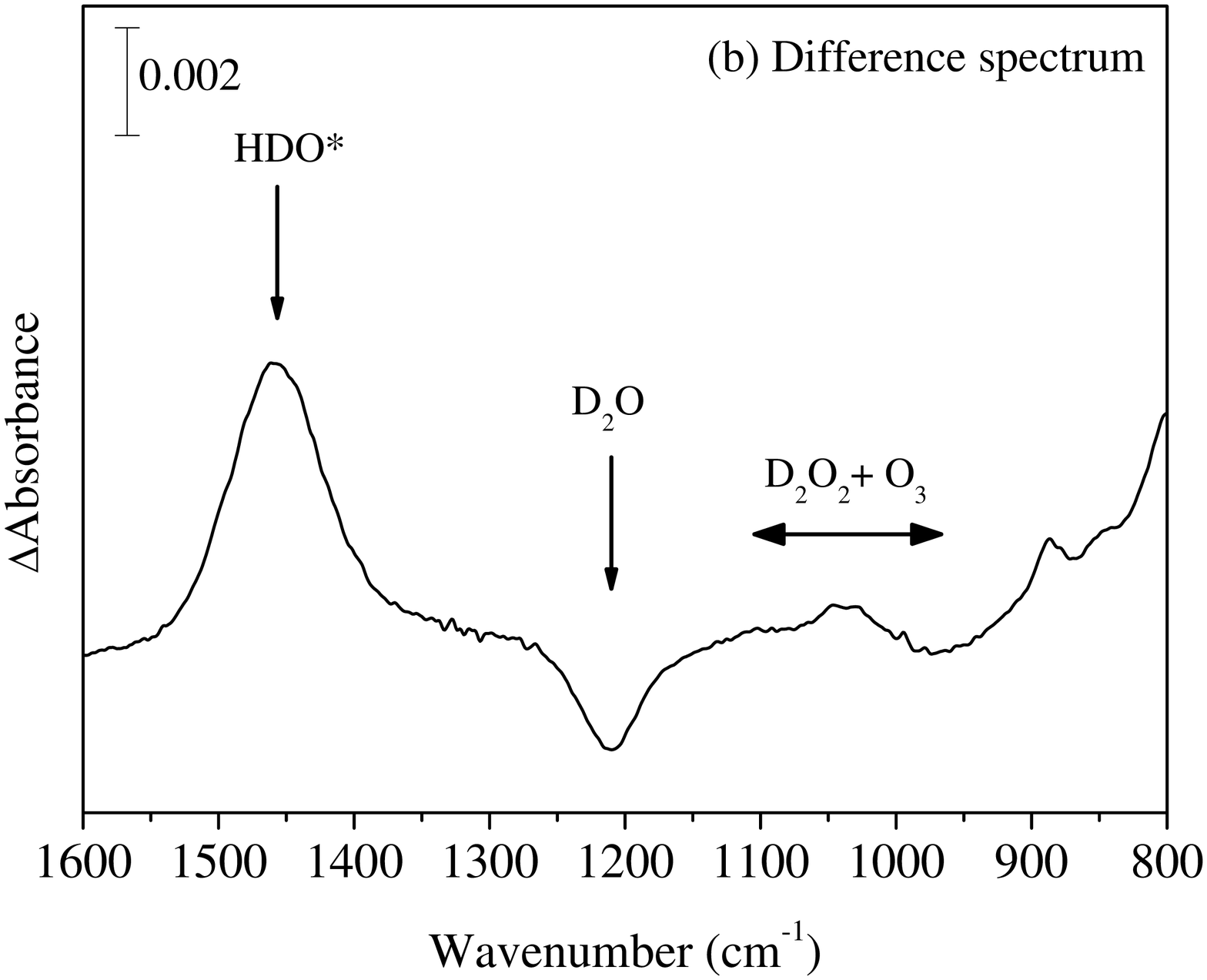}
  \caption{(a) FTIR spectrum after the codeposition of D$_{2}$O fragments with HD and 
(b) the spectral differences relative to the D$_{2}$O-fragment deposition 
sample. The peak with the asterisk (*) was used for quantification purposes.}
  \label{fig8}
\end{figure}

As with the reaction OH + HD, two reaction channels were possible for the reactions of OD with HD:
\begin{eqnarray}
{\rm OD + HD \to HDO + D,}\\
{\rm OD + HD \to D_{2}O + H.}
\end{eqnarray}
For the reasons mentioned above, reaction (\ref{OD+HD->HDO}) was expected to be more efficient than reaction (\ref{OD+HD->D2O}). 
Figure \ref{fig8}(a) shows an IR spectrum of the reaction product obtained after codeposition of D$_{2}$O fragments with HD. 
Several new peaks were observed, most of which corresponded to HDO, as explained earlier. 
The column density of HDO was estimated from the OH-stretching band at 3400 cm$^{-1}$ to be 3.3 $\times $ 10$^{15}$ molecules cm$^{-2}$.

In contrast, the column density of D$_{2}$O did not increase; rather, it decreased upon codeposition, as indicated by the decrease in peak area derived from the OD-bending band for D$_{2}$O at 1200 cm$^{-1}$ (Figure \ref{fig8}(b)). 
This result indicated that reaction (\ref{OD+HD->D2O}) is not effective. 
Assuming that D$_{2}$O was mainly produced through OD + OD $\to $ D$_{2}$O + O in the D$_2$O-fragment deposition, the decrease in the D$_{2}$O column density upon codeposition was mainly attributed to consumption of the corresponding OD by HD through reaction (\ref{OD+HD->HDO}). 
This was based on the fact that reaction (\ref{OD+D2}) (OD + D$_{2}$ $\to $ D$_{2}$O + D) was not efficient, as demonstrated in Section 3.4. 
D$_{2}$O$_{2}$ was assumed to form via the reaction OD + OD, and it may be consumed in the presence of codeposited HD; however, variations in the D$_{2}$O$_{2}$ column density could not be estimated because the OD bending band of D$_{2}$O$_{2}$ at 1000 cm$^{-1}$ mixed with the bands corresponding to other species, such as O$_{3}$ (Figure \ref{fig8}(b)).

\subsection{Relative reaction efficiency}

We found that H$_{2}$O and its isotopologues (HDO and D$_{2}$O) form through surface reactions of OH or OD with H$_{2}$, HD, or D$_{2}$ at 10 K, and we determined the column densities of the reaction products in previous sections (Sections 3.1--3.5). 
The relative efficiency of each reaction was roughly estimated by comparing the column densities of the reaction products obtained under identical conditions (Table \ref{tbl-2}). 
More accurate relative efficiencies, however, could not be obtained due to the reason shown below. 
As indicated in Section 2.1, the accurate fluxes of OH or OD were not measured experimentally, just assigned an upper limit of 1.5 $\times $ 10$^{13}$ radicals cm$^{-2}$ s$^{-1}$, which is equal to the deposition rate of H$_{2}$O or D$_{2}$O when the microwave source is turned off. 
Although we verified by QMS and REMPI that almost all H$_{2}$O and D$_{2}$O molecules were fragmented when the plasma was turned on, unfortunately, the OH flux was not guaranteed to be identical to the OD flux, which makes the determination of the accurate relative efficiencies difficult in the present study. 
Nevertheless, the OH and OD fluxes did not vary significantly from one another because the column densities of the reaction products, such as H$_{2}$O or H$_{2}$O$_{2}$ in the H$_2$O-fragment deposition, were comparable to those of the isotopic counterparts, such as D$_{2}$O and D$_{2}$O$_{2}$, in the D$_2$O-fragment deposition.

The reactions listed in Table \ref{tbl-2} are divided into three groups based on the efficiency of each reaction relative to reaction (\ref{OH+H2}): reactions with comparable efficiency (reactions (\ref{OD+H2}), (\ref{OH+HD->H2O}), and (\ref{OD+HD->HDO})), reactions one order of magnitude less efficient (reactions (\ref{OH+D2}) and (\ref{OD+D2})), and reactions of undetermined efficiency (reactions (\ref{OH+HD->HDO}) and (\ref{OD+HD->D2O})). 
For convenience, hereafter, these three groups are denoted A, B, and C, respectively. 
Although the difference in the barrier height ($\Delta E_{\rm a})$ among the reactions listed in Table \ref{tbl-2} is small \citep[$\Delta E_{\rm a} < \sim $400 K;][]{talu1996,nguy2011}, it is remarkable that a considerable isotope effect was clearly observed in groups A and B. 
The reactions differed intrinsically depending on whether the reaction involved H- or D-atom abstraction, and the effective masses were comparable within a reaction group, as shown in Table \ref{tbl-2}. 
The effective masses for the reactions in group B (0.90) were almost twice as large as those in group A (0.47). 
The inverse correlation between efficiency and the effective mass of reactions is a typical feature of quantum tunneling because the tunneling rate decreases considerably as the effective mass of a reaction increases, as expected from Equation (\ref{tunneling}).

The efficiencies of the reactions in group C can be determined from the increase in the column density of HDO or D$_{2}$O in reaction (\ref{OH+HD->HDO}) or (\ref{OD+HD->D2O}), respectively. 
The efficiency of reaction (\ref{OH+HD->HDO}) could potentially be estimated in a straightforward way by calculating the column density of HDO formed through the reaction. 
This was not possible, however, because the OD-stretching band at 2435 cm$^{-1}$, which was used to calculate the HDO column densities in the present study, included a significant contribution from D$_{2}$O$_{2}$, as explained above. 
The efficiency of reaction (\ref{OD+HD->D2O}) could potentially be estimated from the increase in the D$_{2}$O column density as a result of the reaction. 
The difference between the spectra of the codeposition and the D$_2$O-fragment deposition samples (Figure \ref{fig8}(b)) shows, however, that the D$_{2}$O column density did not increase; rather, it decreased with the codeposition of HD. 
We could not, therefore, calculate the increase in D$_{2}$O column density through reaction (\ref{OD+HD->D2O}). 
The efficiencies of the reactions in group C were thought to be comparable to those in group B because the effective masses are identical, i.e. the reactions in group C were less efficient than those in group A. 
A summary of the discussion in this section is schematically illustrated in Figure \ref{fig9}, which clearly shows that OH and OD preferentially abstract H atom from hydrogen molecules (H$_2$ or HD), leading to H$_2$O and HDO, respectively.

\section{ASTROPHYSICAL IMPLICATIONS}

The present study demonstrated that reactions between OH and H$_{2}$ yield H$_{2}$O on a solid surface at 10 K, despite a large activation barrier (2100 K) for the reaction \citep{atki2004}. 
We assume that reaction (\ref{OH+H2}) occurred in previous experiments on the formation of H$_{2}$O through hydrogenation of O, O$_{2}$, or O$_{3 }$ at 10 K \citep[e.g.][]{miya2008,mokr2009}; however, this assumption was not extensively explored. 
This is the first experimental result to isolate reaction (\ref{OH+H2}) and to show that reactions related to H$_{2}$ actually yielded H$_{2}$O. 
Although in a recent theoretical model, \citet{das2011} assumed that reaction (\ref{OH+H2}) does not occur on grain surfaces in dense clouds because of the activation barrier, the present experimental result strongly suggests that models should include reaction (\ref{OH+H2}) as the efficient route. 

We propose that solid H$_{2}$O in dense molecular clouds is mostly formed by the following three surface reactions: OH + H (reaction (\ref{OH+H})), OH + H$_{2}$ (reaction (\ref{OH+H2})), and H$_{2}$O$_{2}$ + H (reaction (\ref{H2O2+H})). 
The main routes to the formation of H$_2$O in dense clouds are schematically illustrated in Figure \ref{fig10}.
OH would be formed through hydrogenation of O atoms and/or other reactions (e.g., reaction (\ref{H2O2+H})). 
It may be reasonable that H$_{2}$O formation through OH + H$_{2}$ is more efficient than reactions OH + H in dense clouds where H$_{2}$ is more prevalent than H atoms among the various H-related species.
\citet{roma2011} also discussed the importance of reaction (\ref{OH+H2}) relative to reaction (\ref{OH+H}) for the formation of H$_2$O in their experiments where energetic OH was produced by the hydrogenation of O$_3$ at 25--50 K.
As for H$_2$O$_2$, hydrogenation of O$_2$ is probably a unique pathway toward the production of H$_2$O$_2$ in dense clouds.
O$_{2}$ could be produced by the reaction of O + O on grain surfaces and/or supplied from the gas phase. 
The search for solid O$_{2}$ in interstellar environments is ongoing and has not been successful to date. 
Its presence has been suggested in previous studies \citep[e.g.,][]{boog2002,gold2011}. 
The recent detection of H$_{2}$O$_{2}$ in the cloud core $\rho $ Oph A suggested that reaction (\ref{H2O2+H}) may be plausible on grain surfaces \citep{berg2011}. 
The upper limit for the H$_{2}$O$_{2}$ ice abundance relative to H$_{2}$O in several protostars and field stars was estimated based on laboratory IR spectra of H$_{2}$O$_{2}$/H$_{2}$O mixed ice and was found to be 9{\%} $\pm $ 4{\%} \citep{smit2011}. 
In addition, laboratory experiments clearly demonstrated that H$_2$O formation through O$_2$ hydrogenation is efficient under conditions of dense molecular clouds \citep[e.g.,][]{miya2008,oba2009}. 

In addition to the main routes mentioned above, various reactions were proposed that could lead to the formation of H$_{2}$O and other species, such as H$_{2}$O$_{2}$ and OH (Table \ref{tbl-1}). 
These reactions were assumed to make only small contributions to the observed abundance of solid H$_{2}$O in 
dense clouds. 
For example, although hydrogenation of O$_{3}$ was proposed as an important route to the formation of solid H$_{2}$O in dense clouds in previous theoretical \citep{tiel1982} and experimental studies \citep{mokr2009,roma2011}, we suspect that the contribution of this reaction to H$_{2}$O formation is small. 
O$_{3}$ formation is not favored on grain surfaces because the mobility and flux of H atoms is higher than the corresponding properties of O atoms, which prohibits formation of O$_{3}$ by consuming O atoms (reaction (\ref{O+H})) and/or O$_{2}$ (reaction (\ref{O2+H})) prior to O$_{3}$ formation. 
Our conclusion is consistent with the theoretical model proposed by \citet{cupp2007}, who estimated the efficiencies of reactions (\ref{OH+H}), (\ref{H2O2+H}), and (\ref{OH+H2}) in dense clouds at 10 K to be 6{\%}, 17{\%}, and 77{\%}, respectively. 

The formation of solid HDO was experimentally demonstrated to occur in the present study. 
Although HDO in the solid state has been sought after, its detection has not been successful to date. 
An upper limit for the solid HDO/H$_{2}$O ratio in some sources is available \citep{dart2003,pari2003}. 
In the gas phase, the HDO/H$_{2}$O ratios in the low-mass protostars IRAS 16239-2422 and NGC1333-IRAS2A were found to be 0.03 \citep{pari2005} and 0.07 \citep{liu2011}, respectively.

\citet{kris2011} discussed the formation of H$_{2}$O, HDO, and D$_{2}$O through reactions of OH with H$_{2}$, HD, and D$_{2}$, respectively, on ice-covered dust grains in dense molecular clouds. 
They assumed that the reaction OH + HD is barrierless, and proposed that the H$_{2}$O and HDO form statistically in a ratio of 2:1, respectively, from the intermediate H$_{2}$DO. 
Based on these assumptions, this group claimed that the astronomically observed HDO/H$_{2}$O ratio could be explained according to their model; however, the reactions of OH/OD with H$_{2}$/HD/D$_{2}$ do have a barrier, as demonstrated in previous experimental and theoretical studies in the gas phase \citep[e.g.,][]{talu1996,nguy2011}, as well as in the present study. 
Moreover, the present study demonstrated that the formation of HDO through the reactions of OH with HD were much less favored than the formation of H$_{2}$O, which is inconsistent with their assumptions. 
Our results are inconsistent with the theoretical predictions for the formation of H$_{2}$O and its isotopologues by \citet{caza2010,caza2011}, who considered branching ratios of 1 to 1 between reactions (\ref{OH+HD->H2O}) and (\ref{OH+HD->HDO}) or between reactions (\ref{OD+HD->HDO}) and (\ref{OD+HD->D2O}).

\begin{figure}
\plotone{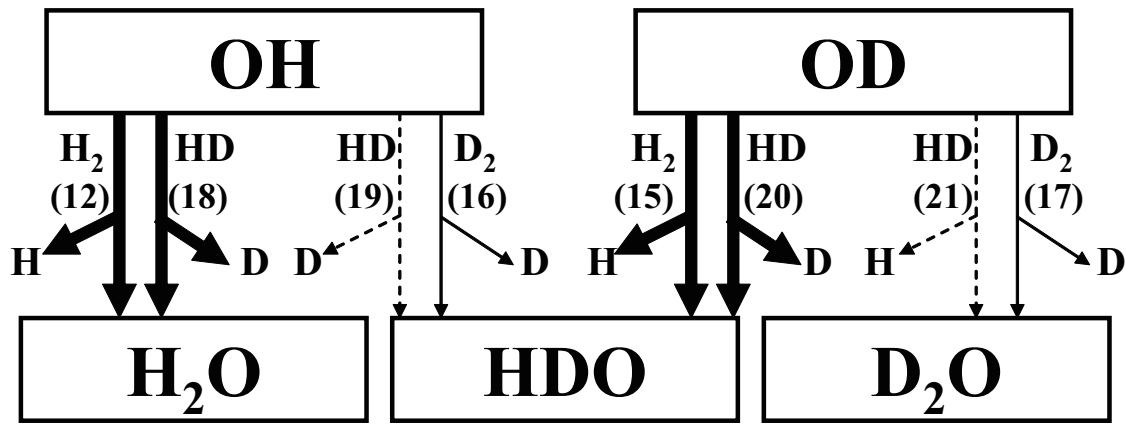} 
  \caption{ Schematic illustration showing the reaction network for the formation of 
solid H$_{2}$O and its isotopologues (HDO and D$_{2}$O) via the reactions 
listed in Table \ref{tbl-2}. The broad or narrow lines represent reactions with 
relative efficiencies of about 1 or 0.1, respectively, compared to reaction 
(\ref{OH+H2}) (Table \ref{tbl-2}). The dashed lines indicate reactions with efficiencies that 
could not be estimated experimentally. The numbers correspond to the 
reaction numbers given in the text and Table \ref{tbl-2}. This figure clearly shows that OH and OD preferentially abstract H atom from hydrogen molecules (H$_2$ or HD), leading to H$_2$O and HDO, respectively.
}
  \label{fig9}
\end{figure}

\begin{figure}
\epsscale{0.5}
\plotone{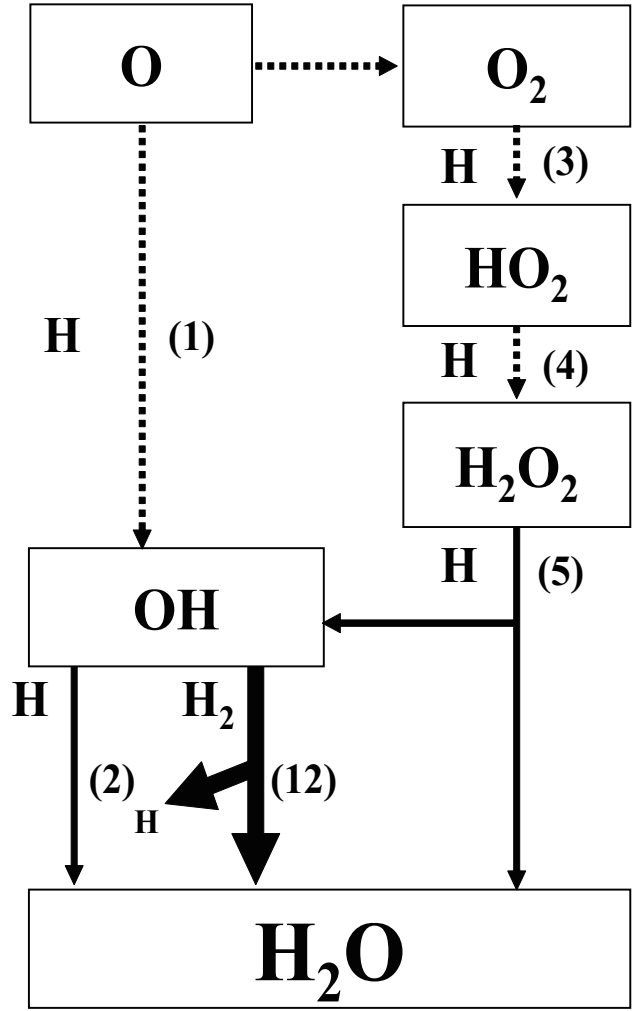} 
\caption{ Main routes to the formation of solid H$_{2}$O in dense molecular clouds: hydrogenation of OH (reaction (\ref{OH+H})), 
the reaction of OH with H$_{2}$ (reaction (\ref{OH+H2})), and the reaction of H$_{2}$O$_{2}$ with H atom (reaction (\ref{H2O2+H})), which are denoted by solid lines.
Other reactions are denoted by dotted lines.
The title reaction OH + H$_2$ is denoted by a thick line.
The numbers in parentheses correspond to the reaction numbers given in the text and Table \ref{tbl-1}.}
  \label{fig10}
\end{figure}

The present experimental results demonstrate that a crucial factor for constraining the ratio of the HDO and H$_{2}$O abundances formed by reactions listed in Table \ref{tbl-2} is the relative abundance of OH and OD in dense clouds. 
The following three reasons support this conclusion: first, the efficiency of H$_{2}$O formation through the reactions of OH with H$_{2}$ (reaction (\ref{OH+H2})) and HD (reaction (\ref{OH+HD->H2O})) is comparable to the efficiency of HDO formation through reactions of OD with H$_{2}$ (reaction (\ref{OD+H2})) or HD (reaction (\ref{OD+HD->HDO})). 
Second, the four reactions (i.e., H-atom abstraction reactions) are much more efficient than the other four reactions 
(i.e., D-atom abstraction reactions) (Table \ref{tbl-2}, Figure \ref{fig9}).
Third, once solid H$_{2}$O is formed, its deuteration does not occur through H--D exchange with D atoms \citep{naga2005} or with D$_{2}$O \citep{galv2011} at 10 K, even if D$_{2}$O ice forms successfully. 
Note that this prediction is only applicable to HDO and H$_{2}$O formation through reactions listed in Table \ref{tbl-2}.

If OH and OD are mainly produced by the reactions of O atoms with H and D atoms, respectively, the relative abundance of H and D atoms would be important to constrain the relative abundance of OH and OD.
According to the deuterium fractionation models in the gas phase proposed by \citet{robe2002}, the atomic D/H ratio is low ($<$ 10$^{-4})$ during the early stages ($<$ 10$^{5}$ years) of molecular cloud evolution. 
After 10$^{5}$ years, the ratio increases dramatically, reaching a value of $\sim$0.06 after 10$^{6}$ years, which is comparable to the observed HDO/H$_{2}$O ratio shown above \citep{pari2005,liu2011}. 
If large regions of an ice mantle are produced during the later stages of molecular cloud evolution, the HDO/H$_{2}$O ratio is expected to reach its highest value for the D/H atomic ratio, $\sim $0.06 \citep{robe2002}. 
If not, other routes may contribute to the observed HDO/H$_{2}$O ratio.

The above scenario is based on an assumption that OH (OD) is produced through the reaction O + H (D) on grain surfaces. 
\citet{robe2002} estimated the OD/OH ratios in their gas-phase model to be 0.093--0.358, higher than the atomic D/H ratio in the same model. 
The participation of gas-phase OD and OH in deuterium fractionation in water formation via surface reactions leads to D enrichment in ice, although the degree of enrichment will depend on the flux of the gas-phase OD and OH onto grains.

Further studies are required to inclusively discuss the formation of solid H$_{2}$O and HDO in dense molecular clouds. 
In future studies, care should be taken when considering chemical reactions, particularly quantum tunneling reactions, with comparable activation barriers because the reaction efficiency can differ dramatically among such reactions, as clearly demonstrated in the present study. 

\acknowledgments

This work was partly supported by a Grant-in-Aid for Scientific Research from the Japan Society for the Promotion of Science (JSPS) and by a research fellowship from JSPS for Young Scientists (Y.O.).




\clearpage




\clearpage

\begin{deluxetable}{clccc}
\tabletypesize{\scriptsize}
\tablecaption{Surface reactions proposed for the formation of solid H$_{2}$O and the related species\label{tbl-1}}
\tablecolumns{5} 
\tablewidth{0pt}
\tablehead{
\colhead{Number} & \colhead{Reaction} & \colhead{{\it E}$_{a}$ (K)} & \colhead{$\Delta${\it H}$^{\circ}$(kJ mol$^{-1}$)} & \colhead{Reference} 
}

\startdata 
1 &  O + H $\to$ \rm OH & 0 & --429 &        \\
2 &  OH + H $\rightarrow$ \rm H$_2$O & 0 & --499 &        \\
3 &  O$_2$ + \rm H $\rightarrow$ \rm HO$_2$ & $\sim${0} & --203 & a      \\
4& HO$_2$ + \rm H $\rightarrow$ \rm H$_2$O$_2$ & 0 & --354  & b       \\
5 & H$_2$O$_2$ + \rm H $\rightarrow$ \rm H$_2$O + OH & 2000 & --285   & c     \\
6 & OH + OH $\rightarrow$ \rm H$_2$O$_2$ & 0 & --211 & d      \\
7 & OH + OH $\rightarrow$ \rm H$_2$O + O & 0 & --71  & d      \\
8 & HO$_2$ + \rm H $\rightarrow$ \rm H$_2$OO$\ast$ $\rightarrow$ \rm H$_2$O + O($^{1}$D) & 0 & --10\tablenotemark{h} & b     \\
9 & HO$_2$ + \rm H $\rightarrow$ \rm H$_2$O..O $\rightarrow$ \rm H$_2$O + O($^{3}$P) & 9300 & --242 & b       \\
10 & HO$_2$ + \rm H $\rightarrow$ \rm H$_2$ + O$_2$($^{3}$$\Sigma$ $_g^{-}$) & 750 & --229 & b       \\
11 & O$_{3}$ + \rm H $\rightarrow$ \rm OH + O$_{2}$ & 500 & --321 & e      \\
12 & OH + H$_{2}$ $\rightarrow$ \rm H$_2$O + H & 2100 & --62  & d       \\
22 & O + H$_{2}$ $\rightarrow$ \rm OH + H & 3160 & +8   & f    \\
23 & HO$_2$ + \rm H$_2 \rightarrow$ \rm H$_2$O$_2$ + H & 13100 & +58 & g      \\
\enddata

\tablenotetext{a}{Walch et al. (1988); Sellevag et al. (2008).}
\tablenotetext{b}{Mousavipour \& Saheb (2007).}
\tablenotetext{c}{Koussa et al. (2006).}
\tablenotetext{d}{Atkinson et al. (2004).}
\tablenotetext{e}{Keyser (1979).}
\tablenotetext{f}{Baulch et al. (1992).}
\tablenotetext{g}{Tsang \& Hampson (1986).}
\tablenotetext{h}{This reaction as a whole is exothermic; however, the latter part of the reaction (H$_2$OO$^\ast \rightarrow$ \rm H$_2$O + O($^{1}$D)) is endothermic by 151 kJ mol$^{-1}$.}
\tablecomments{$\Delta${\it H}$^{\circ}$ denotes the experimental enthalpy change for the gas-phase reactions in kJ mol$^{-1}$, where 1 kJ mol$^{-1}$ corresponds to 120 K. Values of {\it E}$_a$ and $\Delta${\it H}$^{\circ}$ are derived from gas-phase reactions.}
\end{deluxetable}


\clearpage

\begin{deluxetable}{clccccccc}
\tabletypesize{\scriptsize}
\tablecaption{Summary of the Activation Barrier (K), the Effective Mass, and the Measured Efficiency for Reactions OH + H$_2$ and the Isotopologues.\label{tbl-2}}
\tablewidth{0pt}
\tablehead{
\colhead{Number} & \colhead{Reaction} & \colhead{Section\tablenotemark{a}}  &  & \colhead{{\it E}$_a$ (K)} &  & \colhead{Atom} &\colhead{Effective } & \colhead{Relative } \\
&&& \colhead{\tablenotemark{b}} & \colhead{\tablenotemark{c}} & \colhead{\tablenotemark{d}} & \colhead{Abstracted}& \colhead{Mass\tablenotemark{e}} & \colhead{Efficiency\tablenotemark{f}}}
\startdata
12 &OH + H$_2$ $\rightarrow$ \rm H$_2$O + H &3.1 &2000  &2100 &2935 &H  &0.47 &1 \\
15 &OD + H$_2$ $\rightarrow$ \rm HDO + H &3.2 & & &2789  &H  &0.48& $\sim${1}\\
16 &OH + D$_2$ $\rightarrow$ \rm HDO + D &3.3  &2456  & &3026  &D  &0.90 &$\sim${0.1} \\
17 &OD + D$_2$ $\rightarrow$ \rm D$_2$O + D &3.4 & & &2870 &D  &0.91 &$\sim${0.1} \\
18 &OH + HD $\rightarrow$ \rm H$_2$O + D &3.5 &2130 & &2855  &H  &0.48 &$\sim${1} \\
19 &OH + HD $\rightarrow$ \rm HDO + H &3.5  &2130 & &3051 & D&0.90 &n.d.\tablenotemark{g} \\
20 & OD + HD $\rightarrow$ \rm HDO + D &3.5 & & &2703  &H&0.48 &$\sim${1} \\
21 &OD + HD $\rightarrow$ \rm D$_2$O + H &3.5  &  & &2900 &D &0.90 &n.d.\tablenotemark{g} \\

\enddata
\tablenotetext{a}{A corresponding section in the present paper.}
\tablenotetext{b}{Experimentally determined values \citep{talu1996}.}
\tablenotetext{c}{Compilation of literature values \citep{atki2004}.}
\tablenotetext{d}{Computed reaction barrier heights including harmonic zero-point vibrational energy \citep{nguy2011}.}
\tablenotetext{e}{Calculated by using Equation (\ref{mass}) \citep{john1966}.}
\tablenotetext{f}{Estimated by comparison of the yield of the product (H$_2$O, HDO, or D$_2$O) for each reaction with that of H$_2$O for reaction (\ref{OH+H2}).}
\tablenotetext{g}{Not determined due to reasons described in the text.}
\end{deluxetable}






\end{document}